\documentclass[12pt]{article}

\usepackage{amsmath,amsfonts,amssymb}
\usepackage{graphicx}
\usepackage{cite}

\def\plabel#1{\label{#1}}

\newcommand{\be}{\begin{equation}}
\newcommand{\ee}{\end{equation}}
\newcommand{\beq}{\begin{eqnarray}}
\newcommand{\eeq}{\end{eqnarray}}
\newcommand{\bea}[2]{\be\label{#2}\begin{array}{#1}}
\newcommand{\eea}{\end{array}\ee}

\newcommand{\under}[2]{\mathop{#1}\limits_{#2}}

\def\Zb{{\rm \bf Z}}
\def\Sb{{\bf S}}
\def\Ub{{\bf \Upsilon}}
\def\Tb{{\bf \Theta}}

\def\rangl{\right\rangle   }
\def\langl{\left\langle  }
\def\({\left(}
\def\){\right)}
\def\[{\left[}
\def\]{\right]}
\def\p{\partial}

\def\11{1\!\! 1}

\def\hf{{1\over 2}}

\def\eps{\varepsilon}

   \def\CA {{\cal A}}
   \def\CB {{\cal B}}
   \def\CC {{\cal C}}

   \def\CL {{\cal L}}

   \def\CR {{\cal R}}

   \def\CV {{\cal V}}

\def\bz{\bar z}

\def\mul{\mu}
\def\mub{\mu_{_B}}
\def\muc{{\hat\mu}}
\def\mubc{{\hat\mu_{_{B}}}}

\def\ll{l}

\numberwithin{equation}{section}

\DeclareMathOperator{\res}{Res} 

\newcommand{\hw}{\ensuremath{\hat{W}}}

\begin{document}

\thispagestyle{empty}

\begin{flushright}
{\small
ITP-UU-05/16\\
SPIN-05/12\\
hep-th/0504199\\
}
\end{flushright}

\begin{center}

\vspace{1.5cm}

{\LARGE $c=1$ from $c<1$:
Bulk and boundary correlators}\\

\vspace{1cm}

\renewcommand{\thefootnote}{\fnsymbol{footnote}}

{\large Sergei Alexandrov\footnote{email: \texttt{S.Alexandrov@phys.uu.nl}} 
and Emiliano Imeroni\footnote{email: \texttt{E.Imeroni@phys.uu.nl}}}

\setcounter{footnote}{0}
\renewcommand{\thefootnote}{\arabic{footnote}}

\vspace{0.8cm}

\emph{Institute for Theoretical Physics \& Spinoza Institute}

\vspace{0.2cm}

\emph{Utrecht University, Postbus 80.195, 3508 TD Utrecht, The Netherlands}

\vspace{0.4cm}

\end{center}

\vspace{2cm}

\begin{abstract}
We study the $c_L=25$ limit, which corresponds to $c=1$ string theory, of bulk and boundary
correlation functions of Liouville theory with FZZT boundary conditions. This limit is singular 
and requires a renormalization of vertex operators. 
We formulate a regularization procedure which allows to extract finite physical results. 
A particular attention is paid to $c=1$ string theory compactified at the self-dual radius $R=1$.
In this case, the boundary correlation functions
diverge even after the multiplicative renormalization. We show that all infinite
contributions can be interpreted as contact terms arising from degenerate
world sheet configurations. After their subtraction, one gets a well defined set of correlation 
functions. We also obtain several new results for correlation functions in Liouville theory
at generic central charge. 
\end{abstract}

\newpage

\tableofcontents

\section{Introduction}

Non-critical string theories (see for instance~\cite{Klebanov:1991qa,Ginsparg:1993IS,Martinec:2004td}
for a review) have long played the role of an interesting 
laboratory to test various phenomena of critical string theories.
A recent breakthrough
in the understanding of their non-perturbative effects 
\cite{McGreevy:2003kb,Martinec:2003ka,Klebanov:2003km,McGreevy:2003ep,
Alexandrov:2003nn,Douglas:2003up,Gutperle:2003ij,Teschner:2003qk,
Alexandrov:2003un,deBoer:2003hd,Gaiotto:2003yb,Kazakov:2004du,Hanada:2004im,Kutasov:2004fg,
Ambjorn:2004my,Sato:2004tz,Alexandrov:2004cg,
deMelloKoch:2004en,Takayanagi:2005tq,Teschner:2005rd}
drew a lot of attention to these theories.
The ground for this breakthrough was prepared
by a progress in Liouville theory, where
two sets of conformally invariant boundary conditions, 
corresponding to the so called FZZT~\cite{Fateev:2000ik,Teschner:2000md}
and ZZ branes~\cite{Zamolodchikov:2001ah}, were discovered
and all basic boundary correlation functions were calculated 
\cite{Hosomichi:2001xc,Ponsot:2001ng,Ponsot:2003ss}. 
A review of these results can be found in \cite{Nakayama:2004vk}.

Usually, the Liouville correlations functions are found for generic values of the 
Liouville central charge and they are represented as complicated combinations of 
(integrals of) some special functions. On the other hand, the most interesting 
non-critical string theory is the $c=1$ string theory.
Besides giving the first non-trivial example of a non-rational CFT, 
it possesses a clear geometric interpretation as a string theory in
a two-dimensional spacetime. It also allows for non-trivial backgrounds, such as
the two-dimensional black hole, so that it might be useful in the
analysis of black hole thermodynamics. Finally, it appears as a building block in many
constructions of critical superstrings.

The field content of $c=1$ string theory 
consists of a massless scalar matter coupled with the Liouville field with central charge
$c_L=25$. This is a very special value of the central charge for Liouville theory,
since its correlation functions typically diverge at this point. 
This well known divergence is a counterpart of the fact 
that the $c=1$ CFT is a singular 
limit in the family of minimal $(p,q)$ models \cite{Runkel:2001ng}.

In this paper we demonstrate how it is possible to define
all Liouville correlation functions in the $c_L=25$ limit systematically.
The above mentioned divergence at $c_L=25$ can be removed by a 
multiplicative renormalization of the Liouville coupling constants.
Whereas this is a straightforward 
procedure for the theory in the bulk, for the boundary correlation
functions there are additional complications
related to the appearance of the so called ``contact terms'', which
diverge even after the renormalization. In the simplest cases, these contact terms
are entire functions of the boundary cosmological constants and therefore
vanish after the insertion of a finite number of boundary cosmological constant operators.
In the matrix model approach, such terms are usually considered as non-universal
and excluded from the final results, and one can expect the same situation 
to occur in the CFT under consideration.

However, our analysis shows that there are other divergent contributions,
which have a more complicated form and have no simple
analogues in the matrix models. Nevertheless, we show that all of the
divergent terms arising in the $c_L=25$ limit,
independently of their form, can be interpreted as contributions coming from degenerate
two-dimensional geometries, and they should therefore be excluded from 
the resulting expressions for the correlation functions.
In this way one arrives at finite results for all basic correlators, which,
once multiplied by the corresponding matter contributions, 
give the correlation functions of $c=1$ string theory. 
 
Recently, an attempt to extract the $c_L=25$ limit of Liouville theory with boundary 
was made in \cite{Ghoshal:2004mw}. However, the complication related to the appearance 
of the contact terms was not taken into account.
As a result, \cite{Ghoshal:2004mw} provided only divergent results for the leading 
non-universal contributions,
whereas the physical part of the correlation functions remained non calculated.  
Our paper fulfills this gap and presents the physical results derived from a careful 
regularization procedure based on a consistent coupling of Liouville theory and matter CFT. 
As in~\cite{Ghoshal:2004mw}, we limit our analysis to boundary conditions of FZZT type. 
The correlation functions on ZZ branes can be obtained by
applying a well known relation between these two types of 
branes \cite{Martinec:2003ka,Hosomichi:2001xc,Teschner:2003qk,
Ponsot:2003ss,Seiberg:2003nm}.\footnote{Let us note that 
since the results of \cite{Ghoshal:2004mw} 
depend analytically on the boundary cosmological constant, this relation 
implies that the corresponding correlation functions with ZZ boundary 
condition vanish, which is clearly unphysical. In contrast, our results
will ensure non-vanishing correlators for ZZ branes.}

The singularity of the $c_L=25$ limit implies that the final results are simpler 
than those for generic values of the central charge. The greatest simplifications occur
if one restricts oneself to the spectrum of vertex operators corresponding to 
$c=1$ string theory compactified at the self-dual radius $R=1$, which is
the case considered in~\cite{Ghoshal:2004mw}. The reason is that in this case
all operators are degenerate operators in both the matter and Liouville  
sectors. In particular, all special functions reduce to ordinary functions
and all correlators can be given quite explicit expressions.
Therefore, we will be primarily interested in this situation.
The self-dual radius introduces additional 
singularities with respect to the generic case.
We show that, after subtraction of the contact terms, our renormalization ensures that
all correlation functions remain finite.

The restriction to the self-dual radius of compactification is also
relevant for the link to topological strings. It has long been known that $c=1$
non-critical string theory compactified on a circle of radius $R=1$
provides a dual description of topological strings on the conifold
\cite{Imbimbo:1995yv,Ghoshal:1995wm,Dijkgraaf:2003xk,Mukhi:2003sz,Aganagic:2003qj}.
One might therefore expect that the correlation functions that we study here 
shed light on the dynamics of some D-branes in topological string theory.

The organization of the paper is the following.
In section~\ref{secintro} we recall some basic facts about Liouville theory, introduce our 
notations and collect the known results for the correlation functions of Liouville
vertex operators on a sphere and on a disk. In section~\ref{seclimit} we formulate our procedure
to take the $c=1$ limit, based on a CFT representing Liouville theory
coupled to matter with background charge. At this point we also introduce 
the multiplicative renormalization of the vertex operators. 
In the following two sections we compute sphere and disk correlation functions for $c_L=25$.
The former are considered in section 4 and the latter are dealt with in section 5.
Since in the case of boundary correlators the multiplicative renormalization is not enough to remove
all divergences, we begin in subsection~\ref{secuniv} by discussing the
meaning of the contact terms and the necessity of excluding them from
the results. We then proceed with the details of the computation in the remaining part of the section.
Notice that, as an intermediate step in our derivation of the correlators in the $c_L=25$
limit, we also find the Liouville correlation functions
for a generic central charge with momenta corresponding to degenerate matter fields, and
these agree with the predictions of the microscopic approach found in \cite{Kostov:2003uh}.
Finally, we conclude in section~\ref{concl} by summarizing the list of our main results.
Readers who are not interested in the details of the derivations may find this list useful
to get to the results of our work quickly.
Several appendices present various details of our calculations.

\section{Correlation functions in Liouville theory}
\label{secintro}

Liouville theory arises when one considers two-dimensional quantum gravity,
possibly coupled to matter, in the conformal gauge.
It is defined by the following action:
\be
S_L= {1\over 4\pi}\int_{\Sigma}d^2 z\, \sqrt{g} \left( (\p\phi )^2 + Q\hat R\phi
+4\pi\mul \,e^{2b \phi} \right).
\plabel{LIOU}
\ee
It is a CFT whose central charge is given by
\be
c_L=1+6Q^2
\plabel{cliouv}
\ee
and the parameter $b$ in~\eqref{LIOU} is related to Q via the relation
\be
Q=b+{1\over b}.
\plabel{bQ}
\ee
In general, $b$ and $Q$ are determined by the requirement that the
total central charge
of the Liouville and matter field is equal to $26$.
For example, in the case when matter is represented by a
minimal $(p,q)$ model with central charge
$c_{p,q}=1-6{(p-q)^2\over pq}$, the relation \eqref{cliouv} implies that
$b=\sqrt{p/q}$.
Coupling to $c=1$ matter corresponds to the limit $b=1$ so that 
the Liouville central charge becomes $c_L=25$.

\subsection{Sphere correlation functions}

An important class of conformal primaries in Liouville
theory consists of the bulk operators
\be
V_\alpha(z)=e^{2\alpha\phi(z)}\,,
\label{opalph}
\ee
whose scaling dimension is given by $\Delta_\alpha=
\bar\Delta_\alpha=\alpha(Q-\alpha)$.
In particular, the Liouville interaction in the action~\eqref{LIOU} 
can be seen as a perturbation of the Lagrangian by the cosmological constant operator
$\delta\CL=\mul V_b$.

In principle, all information about Liouville theory on a manifold with topology 
of a sphere is contained in the three-point correlation function, which was found to be
\cite{Dorn:1994xn,Zamolodchikov:1995aa}
\be
\langle V_{\alpha_1}(z_1)V_{\alpha_2}(z_2)V_{\alpha_3}(z_3)\rangle_{\rm sphere}=
{C(\alpha_1,\alpha_2,\alpha_3) \over |z_{12}|^{\,2(\Delta_{1}+\Delta_2-\Delta_3)}
|z_{23}|^{\,2(\Delta_{2}+\Delta_3-\Delta_1)} |z_{31}|^{\,2(\Delta_{3}+\Delta_1-\Delta_2)}},
\ee
where
\begin{align}\label{three}
C(\alpha_1,\alpha_2,\alpha_3)&=
\left[\pi\mul \gamma(b^2)b^{2-2b^2}\right]^{\(Q-\sum\alpha_i\)/b} \\
&  \times
{\Ub_{b,0}\Ub_b(2\alpha_1)\Ub_b(2\alpha_2)\Ub_b(2\alpha_3) \over
\Ub_b(\alpha_1+\alpha_2+\alpha_3-Q)\Ub_b(\alpha_1+\alpha_2-\alpha_3)
\Ub_b(\alpha_2+\alpha_3-\alpha_1)\Ub_b(\alpha_3+\alpha_1-\alpha_2)}.\nonumber
\end{align}
Here we used the standard notation for the di-gamma function
\be
\label{ggaamm}
\gamma(x)={\Gamma(x)\over\Gamma(1-x)}\,,
\ee
while the other special functions appearing in \eqref{three} 
and in the following are defined in appendix \ref{A}.

A particular important case of the three-point function is given by the two-point function.
In Liouville theory it is also often interpreted as a reflection amplitude 
which is the coefficient $D_{\rm (refl)}(\alpha)$ appearing after the 
duality transformation $\alpha\to Q-\alpha$:
\be
V_{\alpha}(z)=D_{\rm (refl)}(\alpha)V_{Q-\alpha}(z).
\ee
Its explicit expression reads
\be
D_{\rm (refl)}(\alpha)=\left[\pi\mul \gamma(b^2)\right]^{\(Q-2\alpha\)/b}
{\gamma(2b\alpha-b^2)\over b^2\, \gamma(2-2\alpha/b+1/b^2)}.
\label{tworefl}
\ee

However, in string theory and quantum gravity
the correct way to define the two-point function is through the three-point correlator 
\eqref{three}. As explained for instance in appendix C of~\cite{Aharony:2003vk},%
\footnote{We thank Joerg Teschner for drawing our attention to this reference.}
one should take the three-point function
at the particular values of momenta $\alpha_1=\alpha_2$ and $\alpha_3=b$, and then integrate
it once with respect to $-\mul$. This procedure allows to avoid any problems related
to the residual gauge symmetry on the sphere with two insertions of the vertex operators.
In this way one obtains
\be
D(\alpha)=\left[\pi\mul \gamma(b^2)\right]^{\(Q-2\alpha\)/b}
{(Q-2\alpha)\gamma(2b\alpha-b^2)\over \pi b^{2}\, \gamma(2-2\alpha/b+1/b^2)}.
\label{twosphere}
\ee
Thus, the two-point function found from the three-point correlator differs
from the reflection amplitude by a factor:
\be
D(\alpha)={Q-2\alpha\over \pi}\,D_{\rm (refl)}(\alpha).
\label{difreftwo}
\ee 
In our study we have to use the function \eqref{twosphere}.
In particular, after the renormalization only $D(\alpha)$ gives 
finite expressions for all momenta in the $c=1$ limit.

\subsection{Disk correlation functions}
\label{subsec-cordisk}

When Liouville theory is formulated on a surface with
boundary, there are two possible boundary conditions,
respectively of Neumann and Dirichlet type. The first type gives rise to the so called
FZZT branes~\cite{Fateev:2000ik,Teschner:2000md},
while the second one leads to the ZZ branes~\cite{Zamolodchikov:2001ah}.
In order to work with FZZT boundary conditions,
one adds the following boundary contribution to the action~\eqref{LIOU}:
\be
S_{\rm bnd}={1\over 2\pi }\int_{\p\Sigma} d\xi \,g^{1/4} \(Q\hat K \phi +2\pi\mub e^{b\phi}\).
\ee
The boundary conditions are parameterized by the boundary cosmological constant $\mub$.
However, all correlation functions are more naturally written in terms of another parameter
$s$, which is related to $\mub$ through
\be
{\mub}=\sqrt{\mul\over \sin(\pi b^2)}\cosh(\pi b s) .
\label{fzzsmub}
\ee

In this theory, besides the bulk operators \eqref{opalph}, there are operators acting
on the boundary and changing the boundary conditions
\be
B_\beta^{s_1s_2}(\xi)=e^{\beta\phi(\xi)}.
\label{bndop}
\ee
Here the labels $s_1$ and $s_2$ correspond to the
boundary conditions to the left and right of the point $\xi$ on the boundary.
The scaling dimension of the operator \eqref{bndop} is given by $\Delta_\beta=
\bar\Delta_\beta=\beta(Q-\beta)$.

Let us now summarize the non-trivial correlation functions in the theory with boundary. 
We recall that definitions and properties of the special functions appearing below 
are summarized in appendix~\ref{A}.

\vspace{-0.3cm}\begin{itemize}\setlength{\itemsep}{0pt}

\item \emph{One-point bulk correlation function}~\cite{Fateev:2000ik}
\be
\langle V_{\alpha}(z)\rangle_{\rm disk}={U(\alpha|s)\over |z-\bz|^{2\Delta_{\alpha}}},
\label{onpbulk}
\ee
\be
U(\alpha|s)={1\over \pi b}\,[\pi\mul \gamma(b^2)]^{(Q-2\alpha)/2b}
\Gamma(2b\alpha-b^2)\Gamma(2\alpha/b-1/b^2-1)\cosh\((2\alpha-Q)\pi s\).
\label{onecorL}
\ee

\item {\it Bulk-boundary correlation function}~\cite{Hosomichi:2001xc}
\be
\langle V_{\alpha}(z) B_{\beta}^{ss}(\xi)\rangle_{\rm disk}=
{R(\alpha,\beta|s)\over |z-\bz|^{2\Delta_{\alpha}-\Delta_{\beta}}
|z-\xi|^{2\Delta_{\beta}}},
\label{blkbnd}
\ee
\begin{align}
R(\alpha,\beta|s)&= -2\pi i \left[\pi\mul \gamma(b^2)b^{2-2b^2}\right]^{(Q-2\alpha-\beta)/2b}
{\Gamma_b^3(Q-\beta)\Gamma_b(2\alpha-\beta)\Gamma_b(2Q-2\alpha-\beta) \over
\Gamma_b(Q)\Gamma_b(Q-2\beta)\Gamma_b(\beta)\Gamma_b(Q-2\alpha)\Gamma_b(2\alpha)}
\nonumber \\
&\times\, \int\limits_{-i\infty}^{i\infty}dt\, e^{-2\pi t s}
{\Sb_b(t+\beta/2+\alpha-Q/2)\Sb_b(t+\beta/2-\alpha+Q/2) \over
\Sb_b(t-\beta/2-\alpha+3Q/2)\Sb_b(t-\beta/2+\alpha+Q/2)} .
\label{bulkbnd}
\end{align}

\item {\it Three-point boundary correlation function}~\cite{Ponsot:2001ng}
\be
\langle B_{\beta_3}^{s_1s_3}(\xi_1)B_{\beta_2}^{s_3s_2}(\xi_2)
B_{\beta_1}^{s_2s_1}(\xi_3)\rangle_{\rm disk}=
{C(\beta_1,\beta_2,\beta_3|s_1,s_2,s_3)\over |\xi_{12}|^{\Delta_{1}+\Delta_2-\Delta_3}
|\xi_{23}|^{\Delta_{2}+\Delta_3-\Delta_1} |\xi_{31}|^{\Delta_{3}+\Delta_1-\Delta_2}},
\label{threebnd}
\ee
\beq
& & C(\beta_1,\beta_2,\beta_3|s_1,s_2,s_3)=
-4\pi i
\left[\pi\mul \gamma(b^2)b^{2-2b^2}\right]^{ \(Q-\sum\beta_i\)/2b} 
\nonumber \\
& & \qquad\times\,
{\Gamma_b(\beta_2+\beta_3-\beta_1)\Gamma_b(2Q-\beta_1-\beta_2-\beta_3)
\Gamma_b(Q-\beta_1-\beta_2+\beta_3)\Gamma_b(Q-\beta_1+\beta_2-\beta_3)
\over \Gamma_b(Q)\Gamma_b(Q-2\beta_1)\Gamma_b(Q-2\beta_2)\Gamma_b(Q-2\beta_3)}
\nonumber \\
& & \qquad\times\,
{\Sb_b\(Q-\beta_3+i{s_1-s_3\over 2}\)\Sb_b\(Q-\beta_3-i{s_1+s_3\over 2}\)
\over \Sb_b\(\beta_2+i{s_2-s_3\over 2}\) \Sb_b\(\beta_2-i{s_2+s_3\over 2}\)}
\int\limits_{-i\infty}^{i\infty}dt\, \prod\limits_{i=1}^4 {\Sb_b(t+U_i)\over \Sb_b(t+V_i)},
\label{threepbnd}
\eeq
where
\bea{rclcrcl}{paramet}
U_1&=&Q+i{s_1+s_2\over 2}-\beta_1, & \qquad &
V_1&=&2Q+i{s_2-s_3\over 2}-\beta_1-\beta_3,
\\
U_2&=&Q-i{s_1-s_2\over 2}-\beta_1, & \qquad &
V_2&=&Q+i{s_2-s_3\over 2}-\beta_1+\beta_3,
\\
U_3&=&i{s_2-s_3\over 2}+\beta_2, & \qquad &
V_3&=&Q+is_2,
\\
U_4&=&Q+i{s_2-s_3\over 2}-\beta_2, & \qquad &
V_4&=&Q.
\\
\eea
\end{itemize}

Several comments are in order. 
First of all, one should give a precise definition of
the contours of integration in~\eqref{bulkbnd} and~\eqref{threepbnd}. 
The rule is the following: for both integrals, 
the integration contour in the complex $t$-plane 
lies on the right of the poles of the integrand coming from poles of the numerator 
and to the left of the poles coming from zeros of the denominator.
This gives an unambiguous definition of the integrals, except in the case when a pole from 
the numerator coincides with a pole from the denominator. If such a situation of colliding poles 
occurs, it means that the integral has a singularity at the corresponding 
point in the parameter space. 

Another comment is that we have changed the overall normalization 
of the one-point bulk and the three-point
boundary correlation functions with respect to the one usually found in the literature.
Compared to the standard expressions from~\cite{Fateev:2000ik}
and~\cite{Ponsot:2001ng}, the first one is divided by $2\pi$
and the second one is multiplied by $4\pi$.
This is done because the theory is completely defined by
the bulk-boundary and the three-point correlation functions, whereas
the one-point bulk correlator is a derived quantity,
whose normalization can be fixed starting from any of the two. For example, 
one can take the bulk-boundary correlation function~\eqref{bulkbnd} with
the boundary operator being the boundary cosmological constant operator, {\it i.e.} $\beta=b$,
and then integrate with respect to $-\mub$. In appendix \ref{B},
we show that this procedure gives precisely \eqref{onecorL} with the factor $1/2\pi$ included.
This factor can be traced back to the rotational symmetry of the disk, which remains non-fixed
in the one-point correlator. The correct normalization valid for string theory
requires to divide by the volume of the gauge group and thus follows from the bulk-boundary
correlation function where all symmetries are fixed. 

However, a similar procedure (see below) done for the standard expression for the three-point 
function leads to a result differing
by a factor $4\pi$. 
This means that there is a mismatch between 
the bulk-boundary and the three-point functions. 
Assuming that the bulk-boundary correlator is correct, 
one should then modify the standard expression for the three-point correlator 
by including this factor, as we did in~\eqref{threepbnd}.

The third comment is that, similarly to the bulk case, 
one can also consider the two-point boundary correlation function.
Its usual definition used in Liouville theory is through the reflection relation \cite{Fateev:2000ik}
\be
B_\beta^{s_1s_2}(\xi) =d_{\rm (refl)}(\beta|s_1,s_2)\, B_{Q-\beta}^{s_1s_2}(\xi),
\label{twobnd}
\ee
where
\be
d_{\rm (refl)}(\beta|s_1,s_2)={\left[\pi\mul \gamma(b^2)b^{2-2b^2}\right]^{(Q-2\beta)/2b}
\Gamma_b(2\beta-Q)\Gamma_b^{-1}(Q-2\beta)
\over \Sb_b\(\beta+i{s_1+s_2\over 2}\)\Sb_b\(\beta-i{s_1+s_2\over 2}\)
\Sb_b\(\beta+i{s_1-s_2\over 2}\)\Sb_b\(\beta-i{s_1-s_2\over 2}\) }.
\label{twopbnd}
\ee
However, as usual, the two-point function relevant for quantum gravity should be
obtained through the three-point function \eqref{threepbnd} by the standard procedure
of taking one of the operators to be the boundary cosmological 
constant operator and integrating with respect to $-\mub$. In fact, since the initial expression is not 
explicitly symmetric, there are three inequivalent ways to do this depending on which
operator we choose to represent $B_b^{ss}$. 
In appendix \ref{C} we show that all three ways lead to the same result. This fact provides
a non-trivial explicit check of the symmetry of the three-point function \eqref{threepbnd}
under cyclic permutations of its arguments. 
We find that, as in the bulk case \eqref{difreftwo}, the two-point function
differs from the reflection amplitude \eqref{twopbnd} by a simple (momentum dependent) factor 
\be
d(\beta|s_1,s_2)={2(Q-2\beta)}\, d_{\rm (refl)}(\beta|s_1,s_2).
\label{twofunrel}
\ee

In appendix~\ref{ap-onepoint}, starting from the two-point function found in this way,
we also rederive the one-point function
of the cosmological constant operator. The result should coincide with the one obtained
from the bulk-boundary correlator, and this forced us to introduce the $4\pi$ factor
mentioned above. These results also contain information about the one-point function
of the boundary cosmological constant operator which is found to be 
\be
W(s)={2b\, \over \pi}\left[\pi\mul \gamma(b^2)\right]^{{1\over 2b^2}}
\Gamma\(2-{1\over b^2}\)\, \cosh(\pi s /b).
\label{oneW}
\ee
This quantity will be quite important in the following so it is useful to give its explicit
expression at this point.

\section{The $c=1$ limit}
\label{seclimit}

\subsection{2D string theory with a background charge}

It is well known that the $c=1$ limit of the family of CFTs with central charge
$c\le 1$ coupled to Liouville theory with $c_L=26-c$ is singular. 
Therefore, one cannot put directly $b=1$ in the results obtained
in Liouville theory with a generic central charge. To make the limit meaningful, 
we perform a regularization by
considering Liouville theory coupled to a Gaussian matter field
with a background charge $q$, following~\cite{Kostov:2003uh,Kostov:2003cy}.
The Gaussian field is described by the following action 
\be
S_{\rm mat}={1\over 4\pi}\int_{\Sigma}d^2z\, \sqrt{g}\( (\p x)^2+iq x\hat R\)+
{1\over 2\pi}\int_{\p\Sigma}d\xi g^{1/4} \(iqx\hat K \).
\label{matact}
\ee
The background charge leads to a shift of the central charge
of the matter field
\be
c_{\rm mat}=1-6q^2.
\label{matcc}
\ee
Requiring that the total central charge of the system vanishes
\be
c_{\rm tot}=c_L+c_{\rm mat}+c_{\rm ghost}=\(1+6Q^2\)+\(1-6q^2\)-26=0,
\label{ctot}
\ee
one can express the background charge in terms of the Liouville parameter $b$
\be
q={1\over b}-b.
\label{cb}
\ee
The bulk and boundary vertex operators of the full theory
\be
\CV_p=e^{-ipx}e^{2\alpha(p)\phi}, \qquad
\CB_p=e^{-ipx}e^{\beta(p)\phi}
\label{verop}
\ee
are found from the condition that their dimensions are equal to 1
\beq
\Delta\[\CV_p\]&=&{p\over 2}\({p\over 2}+q\)+{\alpha}\(Q-{\alpha}\)=1,
\\
\Delta\[\CB_p\]&=&{p}\(p+q\)+{\beta}\(Q-{\beta}\)=1.
\label{conddim}
\eeq
Taking into account the Seiberg bound~\cite{Seiberg:1990eb}, 
this gives for the parameters $\alpha$ and $\beta$
\bea{llll}{parverp}
\alpha(p)=b- p/2 & {\rm \ if \ \ }p>-q, 
& \quad \alpha(p)=1/b+p/2 &{\rm \ if \ \ }p<-q,
\\
\beta(p)=b-p & {\rm \ if \ \ }p>-q/2,
& \quad \beta(p)=1/b+p  & {\rm \ if \ \ }p<-q/2.
\eea
The two branches of the solution correspond to the left and right moving 
(or positive and negative chirality) tachyons
in the target space picture of two-dimensional string theory.

In order to restrict ourselves to the spectrum of the theory compactified at radius $R$,
we have to take the discrete set of operators with quantized momenta
\be
p_n={n/R}.
\label{spectp}
\ee
Then for $n\ge 0$ the corresponding Liouville momenta are 
\be
\alpha_n=b-{n\over 2R}, \qquad \beta_n =b-{n\over R}.
\label{alphbet}
\ee
In fact, for generic $R$ and $n\ne 0$, the correlation functions do not  
exhibit singularities. 
Therefore, the details of the limiting procedure are not important in this situation.

However, if $R$ is rational some additional singularities appear so that the $c=1$ limit
should be treated with a great care.
We will pay special attention to the case of the self-dual radius $R=1$.
On the one hand, it is the most singular case, but on the other hand
it allows the most explicit representation for the correlation functions.
The origin of the additional divergences can be traced back to the fact 
that all operators with momenta $p_n=n$
are Liouville dressed degenerate matter fields of the $c=1$ CFT.
Therefore, it is natural to take the $c=1$ limit in such a way that the operators
in question are degenerate for all $b$.
For general $b$ the momenta corresponding to the bulk degenerate fields are
\be
p_{r,s}^{\rm bulk}=-q + r/b- sb,
\qquad  
\alpha_{r,s}=Q/2-\left|{r\over 2b}-{sb\over 2}\right|,
\label{degblk}
\ee
and those of the boundary degenerate fields are
\be
p_{r,s}^{\rm bnd}=\(-q + r/b- sb\)/2, 
\qquad 
\beta_{r,s} =Q/2-\left|{r\over 2b}-{sb\over 2}\right|,
\label{degbnd}
\ee
where in both cases $r,s\in \Zb,\ rs \ge 0$.
Thus, when dealing with the case $R=1$, in the regularized theory we will set $R=b$  
so that $p_n=n/b$ and the Liouville momenta \eqref{alphbet} take the form
\be
\alpha_n=b-{n\over 2b}, \qquad \beta_n =b-{n\over b}, \qquad n\ge 0.
\label{alphbeto}
\ee
This corresponds to the choice $r=n+1,\ s=1$ for the bulk and $r=2n+1,\ s=1$
for the boundary momenta.
 
However for $n<0$ this solution does not work. To get degenerate fields 
with momenta of the form \eqref{spectp}, one has to choose $R=1/b$ and,
for example, in the bulk case $r=1,\ s=1-n$ so that $p_n=nb$.
This reflects difficulties in compactifying a field with a background charge.
After compactification, there are degenerate fields only of either 
positive or negative chirality. On the other hand,
to write non-vanishing correlation functions, one needs to satisfy 
the momentum conservation law in the matter sector, which due to the presence 
of the background charge $q$ acquires an additional term.
For the sphere and disk correlation functions it takes respectively the following form
\be
\sum\limits_{i=1}^k p^{\rm bulk}_i=-2q, \qquad 
\sum\limits_{i=1}^k p^{\rm bulk}_i+\sum\limits_{i=1}^l p^{\rm bnd}_i=-q.
\label{momcon}
\ee
This implies that at least one momentum must have $r,s\le 0$ and hence
must not be of the form \eqref{spectp} required by compactification. 
Thus, strictly speaking, we have a compactified theory only in the case $b=1$. 

Usually, there are several ways to satisfy the conservation condition \eqref{momcon}
which differ by the choice of chiralities of the vertex operators. In the $b=1$ limit
they correspond just to different choices of signs of momenta
and should lead to the same results.
We are then free to choose a simple convention consisting in taking all operators except one 
to be of positive chirality. In addition, on the disk we will always choose
a boundary operator to be the one with negative chirality.
Then, if for positive chirality $p^{\rm bulk}_i=n_i/b$ and $p^{\rm bnd}_i=m_i/b$,
the remaining momentum must be of the form 
\be
p^{\rm bulk}_{k}=-2q-n/b, \qquad 
p^{\rm bnd}_{l}=-q-m/b,
\label{pnnn}
\ee 
where $n=\sum_{i=1}^{k-1}n_i$ and $m=\sum_{i=1}^{k}n_i+\sum_{i=1}^{l-1}m_i$,
respectively on the sphere and on the disk.
The momenta \eqref{pnnn} are also degenerate and 
can be obtained from \eqref{degblk} and \eqref{degbnd} for 
$r=-(n+1),\ s=-1$ and $r=-(2m+1),\ s=-1$, respectively. 
Starting from the momenta \eqref{pnnn} for the matter part,
one arrives at the same Liouville momenta as in \eqref{alphbeto}.

We conclude that, after fixing all Liouville momenta to be of the form \eqref{alphbeto},
the only remaining free parameter in our system is $b$.
We are interested in the limit $b\to 1$ where we should find the compactified
$c=1$ string theory. In order to take this limit, we put $b=1-\eps$ for a small parameter $\eps$ and
extract all singular and finite terms in the correlation functions.
We will explain that the singular contributions are given by the so called contact terms,
which are to be neglected, while the physical answers are only contained in the finite
part of the correlation functions.
The exact meaning of the contact terms will be explained in section \ref{secuniv}.

\subsection{Renormalization of couplings and vertex operators}
\label{s:renorm}

It is known that to get a finite sensible $c=1$ limit, one should
simultaneously renormalize the couplings of the theory 
(see for example~\cite{Martinec:2003ka,Alexandrov:2004ks}). The basic
Liouville couplings are renormalized as follows
\be
\muc=
\pi\gamma(b^2)\mul,
\qquad
\mubc=
{\pi\mub\over\Gamma(1-b^2)}.
\label{mucone}
\ee
The normalization is chosen so that the parameterization of
the renormalized boundary cosmological constant looks simply as
\be
\mubc=\sqrt{\muc}\cosh(\tau),
\label{mubmuc}
\ee
where we introduced a new parameter 
\be
\tau=\pi bs .
\label{taus}
\ee
The introduction of the parameter $\tau$ is very convenient. It allows to
avoid complications related to the fact that,
although in the $c=1$ limit the parameter $b$ disappears from~\eqref{taus},
it is important to keep track of it to correctly derive the
physical part of the correlation functions in the boundary theory.
Another useful notation, which we will extensively use, is
\be
\hat\mu_{ij}=\mubc(\tau_i)-\mubc(\tau_j).
\label{notmu}
\ee

The definitions \eqref{mucone} imply the renormalization of the bulk 
and boundary cosmological constant operators.
We will show that, in order to get finite correlation functions at $R=1$,
one should renormalize also all other vertex operators.
Generalizing \eqref{mucone} and following \cite{Kostov:2003uh,Kostov:2003cy,Kostov:2005kk},
where the multiplicative renormalization was used to compare the CFT results 
with results of the discrete approach to 2D quantum gravity, 
we define the renormalized bulk and 
boundary operators with momentum $p$ as follows
\be
\hat\CV_p={1\over \pi\gamma(\alpha^2(p)-p^2/4)}\, \CV_p,
\qquad 
\hat\CB_p={\Gamma(1-\beta^2(p)+p^2)\over \pi}\,\CB_p,
\label{tone}
\ee
where the Liouville momenta $\alpha(p)$ and $\beta(p)$ were defined in \eqref{parverp}.
Their explicit expressions depend on the sign of $p$ and thus operators of positive
and negative chirality have different renormalization factors:
\be
\hat\CV_p^+={1\over \pi\gamma(b^2-bp)}\, \CV_p^+
,
\qquad 
\hat\CB_p^+={\Gamma(1-b^2+2bp)\over \pi}\,\CB_p^+
,
\label{tonep}
\ee
and 
\be
\hat\CV_p^-={1\over \pi\gamma(1/b^2+p/b)}\, \CV_p^-
,
\qquad 
\hat\CB_p^-={\Gamma(1-1/b^2-2p/b)\over \pi}\,\CB_p^-
.
\label{tonem}
\ee
Notice that the renormalization factors for different chiralities become the same in the $b=1$
limit when expressed as functions of Liouville momenta.

The renormalized correlation functions are defined in terms of vertex operators 
$\hat\CV_p$ and $\hat\CB_p$
\be
\hat\CA\(\{p_i^{\rm bulk}\},\{p_j^{\rm bnd}\}|\{s_j\}\)
=
\langl\prod\limits_i  \hat\CV_{p_i^{\rm bulk}}(z_i) \prod\limits_j  
\hat\CB_{p_j^{\rm bnd}}^{\,s_{j+1} s_{j}}(\xi_j)\rangl 
\label{corlim}
\ee
and can be obtained from the non-renormalized expressions by multiplying them with 
the factors in~\eqref{tone}.
The correlation functions of the $c=1$ theory are found then
in the limit $b\to 1$ 
\be
\CA_{\,c=1}\(\{p_i^{\rm bulk}\},\{p_j^{\rm bnd}\}|\{s_j\}\)
=\under{\lim}{b\to 1}\hat\CA\(\{p_i^{\rm bulk}\},\{p_j^{\rm bnd}\}|\{s_j\}\) .
\label{limcorf}
\ee 
Actually, in this paper we will consider only the Liouville part of 
the correlation functions. However, especially in the boundary case, this is
the main non-trivial part which contains
all information about the dependence on the boundary cosmological constant.

It is clear that in the limit $b\to 1$ the renormalization factors of both bulk
and boundary cosmological constants appearing in \eqref{mucone} vanish, whereas for $p_n$
with $n>0$ (see \eqref{tonep}) the corresponding factors diverge as ${\eps^{-1}}$
in the bulk case and remain finite for the couplings of the boundary operators.
The necessity of a renormalization with
such properties can be understood as follows.
First, the analysis of sphere correlation functions fixes
the renormalization of all bulk operators, which can also be confirmed
by the fact that it gives a finite result for the
one-point bulk correlator on a disk.
Then, the renormalization of the boundary operators follows from the study
of the bulk-boundary function and can be verified by analyzing the boundary 
two- and three-point functions.
In the next sections we will study in detail all
basic correlation functions of Liouville theory following this sequence.

\section{Sphere correlation functions in the $c=1$ limit}

First, we evaluate the $c_L=25$ limit of the bulk correlation functions 
for Liouville theory on a sphere.
The main quantity to consider is the three-point function \eqref{three}.
For the momenta \eqref{alphbet} it takes the form
\beq\label{threerR}
 C(\alpha_{n_1},\alpha_{n_2},\alpha_{n_3})&=&
\left[\pi\mul \gamma(b^2)b^{2-2b^2}\right]^{{1\over b^2}+{1\over 2b R}\sum n_i-2}\\
& & \hspace{-2cm}
 \times{\Ub_b(b)\Ub_b\(2b-{n_1\over R}\)\Ub_b\(2b-{n_2\over R}\)\Ub_b\(2b-{n_3\over R}\) \over
\Ub_b\(2b-{1\over b}-{n_1+n_2+n_3\over 2R}\)\Ub_b\(b+{n_3-n_1-n_2\over 2R}\)
\Ub_b\(b+{n_1-n_2-n_3 \over 2R}\)\Ub_b\(b+{n_2-n_3-n_1 \over 2R}\)} .\nonumber 
\eeq
This expression can be simplified if one takes into account 
the momentum conservation \eqref{momcon} from the matter sector,
which implies that $n_3 = n_1+n_2$. 
We also apply here the multiplicative renormalization~\eqref{tone}.
As explained in the previous section, our convention is to
choose two operators of positive chirality
and one of negative chirality, which in this case has to be
the third one due to the just mentioned condition on $n_i$. Thus, the expression~\eqref{threerR} 
must be multiplied by two factors from~\eqref{tonep} and one from~\eqref{tonem}. 
As a result, it reduces to
\be
\hat C(\alpha_{n_1},\alpha_{n_2},\alpha_{n_1+n_2})=
{\muc^{{1\over b^2}+{n_1+n_2\over bR}-2} \over \pi^3 b}.
\label{threeconR}
\ee
The $b=1$ limit of this three-point function is well defined and gives
\be
C_{c=1}(\alpha_{n_1},\alpha_{n_2},\alpha_{n_1+n_2})
={ 1\over \pi^3 }\,\muc^{{n_1+n_2 \over R}-1}.
\label{threerenR}
\ee
One can show that for other choices of chiralities and for generic central charge the result
differs from \eqref{threeconR} only by a power of $b$ 
and hence it leads to the same expression \eqref{threerenR} in the $c=1$ limit.
This shows that our definition of the $c=1$ limit does not depend on the artifact
of the choice of chiralities.

As explained above, the two-point function is obtained from the three-point correlator
by putting one of the Liouville momenta equal to $b$ ($n=0$) 
and integrating with respect to (minus) the cosmological constant. 
For \eqref{threerenR} this procedure gives
\be
D_{c=1}(\alpha_0)=-{1\over \pi^3 }\,\log\muc,
\qquad \ \
D_{c=1}(\alpha_n)=-{R\over \pi^3 n}\,\muc^{n\over R}, \quad n\ge 1.
\label{tworen}
\ee
Of course, the same results can be found from the explicit 
expression \eqref{twosphere} for the two-point function at generic central charge.

Notice also that, if we had used the reflection amplitude $D_{\text{(refl)}}(\alpha)$ 
given in~\eqref{tworefl}
as a starting point, instead of the two-point function derived from the three-point correlator,
the computation in this subsection would have yielded a divergent result for
$D_{c=1}(\alpha_0)$. This provides a confirmation of the fact that the two-point function that
we use is the one which is relevant for $c=1$ string theory.

To conclude, we see that the renormalization~\eqref{tone} is sufficient
to produce finite results for sphere correlators in the $b= 1$ limit.
The result is analytic in the compactification radius $R$
and there are no additional singularities arising at the self-dual point.

\section{Disk correlation functions in the $c=1$ limit}
\label{secdisk}

\subsection{Divergent contributions as contact terms}
\label{secuniv}

Let us now turn to the correlation functions characterizing the theory on the disk
with FZZT boundary conditions. In our analysis we will encounter a new phenomenon.
It turns out that all correlation functions in the boundary theory at
the self-dual compactification radius $R=1$, even after
the multiplicative renormalization~\eqref{tone} 
of the bulk and boundary vertex operators, are divergent
in the $c=1$ limit. However, we are now going to argue that
the divergent contributions have the interpretation of \emph{contact terms}, namely
degenerate world sheet configurations,
and should be omitted from the physical answers.\footnote{The picture presented here
arose in discussions with Ivan Kostov.} 
In order to understand the nature of the divergences, 
it is useful to anticipate here the general structure 
of the correlation functions in the $c=1$ limit, which will be 
studied in full detail in the following subsections. 

First, it is very easy to characterize the correlation functions
of operators with only vanishing boundary momenta of the matter part.
As we will see, in this case the leading term diverges as $1/\eps$ and 
is an entire (polynomial) function of the boundary cosmological constants. 
Such terms are well known in the matrix approach
to non-critical strings where they are called {\it non-universal contributions} and 
usually neglected. This is justified by the fact that they vanish 
after a finite number of differentiations with respect to $\mub$,
which means that such contributions disappear from the correlation functions 
with enough number of insertions of the boundary cosmological constant operators.
In addition, since such terms are entire functions of $\mub$, they do not contribute 
to correlation functions with ZZ boundary conditions, which can be obtained 
from non-trivial monodromy properties of correlators with FZZT conditions
\cite{Teschner:2003qk,Seiberg:2003nm,Alexandrov:2004ks}.

\begin{figure}
\begin{center}
\includegraphics[scale=0.5]{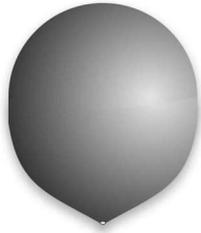}
\caption{{\small Degenerate world sheet configuration with shrinking boundary.
It looks like a sphere with a puncture appearing at the place of the degenerating hole.}}
\label{f:zerolength}
\end{center}
\end{figure}

The meaning of these non-universal contributions 
becomes clear when one passes to the {\it length representation}
of correlation functions~\cite{Moore:1991ir,Moore:1991ag}.
Suppose we have a correlation function $C(\mub)$
on the disk with FZZT boundary conditions described by the boundary cosmological 
constant $\mub$. The parameter $\mub$ is conjugated to the length of the boundary, 
and the correlation function can then be thought as a Laplace transform
of the corresponding correlation function with a fixed boundary length 
\be
C(\mub)=\int\limits_0^{\infty}d\ell\,e^{-\,\ell\,\mub}\, \tilde C(\ell)\,.
\label{lengmub}
\ee
Let us concentrate on a contribution to $C(\mub)$ which is polynomial in $\mub$.
Applying the inverse Laplace transform, one finds that it contributes to
$\tilde C(\ell)$ only at $\ell=0$:
\be
\int\limits_{-i\infty}^{i\infty}d\mub\,e^{-\,\ell\,\mub} \, P(\mub)
\sim P(-\p/\p \ell)\,\delta(\ell)\,. 
\label{divconC}
\ee
Thus, the contributions which are entire functions of the boundary cosmological 
constant correspond to contributions of degenerate world sheets with boundaries
shrunken to a point, as shown in fig.~\ref{f:zerolength}. The same conclusion remains true
if different segments of the boundary are characterized by different
boundary cosmological constants. Such degenerate configurations 
should be excluded from the sum over two-dimensional geometries and, correspondingly,
from the final result.\footnote{These degenerate contributions may in fact still contain
some physical information. Moreover, in some cases they do represent the physical part of the 
correlation functions as it happens, for example, in the case of topological gravity 
corresponding to $c=-2$ \cite{Gaiotto:2003yb}. However, the $c=-2$ case is special
due to the existence of two different theories with the same matter central charge:
the above mentioned topological gravity and the so called bosonic
string embedded in $-2$ dimensions~\cite{Kostov:1991cg}.
Their correlation functions can be obtained either as finite parts of
the limits $b^2\to 2$ and $b^2\to 1/2$ respectively, or as infinite and finite parts of 
the latter limit. However, we believe that two-dimensional or $c=1$ string theory is 
defined unambiguously and its boundary correlators are given by finite non-degenerate
contributions.}

The case when a correlation function involves boundary operators with non-vanishing
boundary momenta is more complicated. In this case there are divergent terms
up to order $\eps^{-n_{\rm tot}-1}$, where $n_{\rm tot}=\sum n_i$ 
is the total momentum of the boundary operators. 
The leading divergent terms are still polynomials in the boundary cosmological 
constants. It is these leading contributions that were calculated in \cite{Ghoshal:2004mw}, 
although our answers even for these terms are different from those of \cite{Ghoshal:2004mw}.
The reason for this discrepancy is that the $c=1$ limit requires to be carefully 
defined and cannot be taken without considering the consistency of 
the coupling between the matter part and the Liouville part.

\begin{figure}
\begin{center}
\includegraphics[scale=0.72]{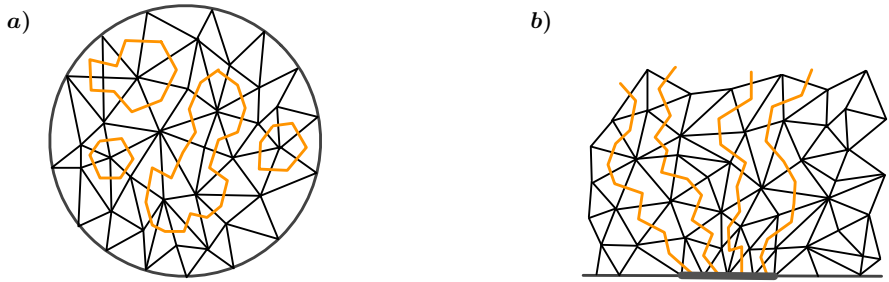}
\caption{{\small The loop gas representation of correlation functions in the SOS model.
a) A typical contribution to the sum over triangulations. b) The loop gas representation 
of a boundary operator ($n=2$). In the continuum limit all lines come from the same single point of 
the boundary.}}
\label{f:loopgas}
\end{center}
\end{figure}
Subleading divergent terms are not entire functions anymore. Therefore,
the previous reasoning based on~\eqref{divconC}, which allowed us to neglect
the divergent contributions, cannot be directly applied and some
additional argumentation is needed.
A key observation is that the boundary correlation functions 
with momenta~\eqref{alphbeto} can be written as polynomials in the one-point 
function $W(s)$ of the boundary cosmological constant operator, which was given in \eqref{oneW}.
This fact implies that all divergent terms come from the divergences of 
this function, and therefore one can remove all divergent contributions by renormalizing $W(s)$. 

To understand the origin of this renormalization, it is useful to
consider how these correlation functions are represented in the discrete approach to 
2D quantum gravity. In this approach the system we are dealing with,
namely Liouville theory coupled with a matter CFT, can be realized as the so called
solid-on-solid (SOS) model on a fluctuating lattice.\footnote{The corresponding 
matrix model realization was constructed in \cite{Kostov:1992ie}.}  
We are not going to present the SOS model here, and we refer
the reader to \cite{Kostov:2003uh} for the relevant details.  
What is important for us is the loop representation of this model 
\cite{Kostov:1989eg,Kostov:1991cg,Kazakov:1991pt}, according to which the matter fields 
can be seen as non-intersecting loops on randomly triangulated surfaces.
Correlation functions are given by a sum over all triangulations and all
admissible configurations of the loops.  A typical configuration contributing 
to the sum is shown in fig.~\ref{f:loopgas}a.

The degenerate boundary operators with momentum $p_n=n/b$ in the matter part 
are realized in the loop representation by the so called ``star'' operators,
that create $2n$ lines at some point of the boundary (see fig.~\ref{f:loopgas}b).
These lines are also non-intersecting and end at another point 
of the boundary. This fact allows to represent the general structure
of the boundary correlation functions and was the starting point 
for their derivation from the discrete approach \cite{Kostov:2002uq,Kostov:2003uh}.  

\begin{figure}
\begin{center}
\includegraphics[scale=0.8]{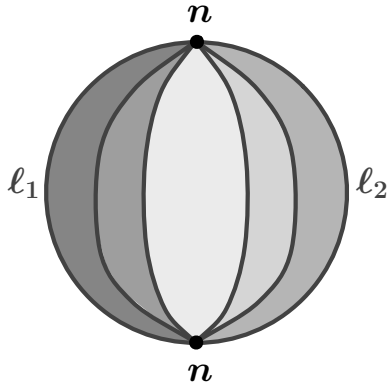}
\caption{{\small The two-point function $\tilde d_n(\ell_1,\ell_2)$.}}
\label{f:2point}
\end{center}
\end{figure}
\begin{figure}
\begin{center}
\includegraphics[scale=0.8]{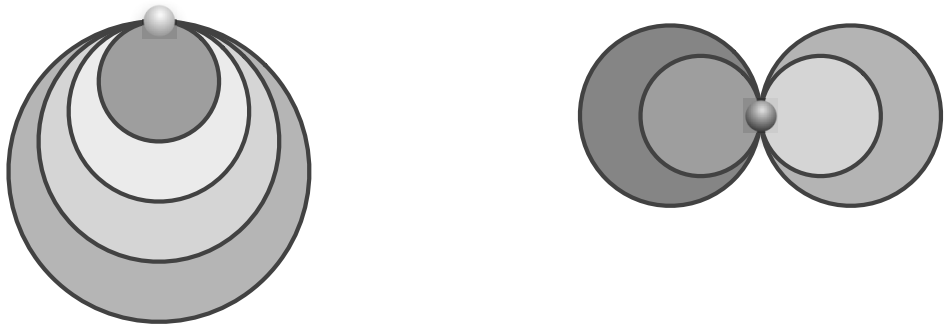}
\caption{{\small Possible degenerations of the two-point function. The first picture shows 
the case when the boundary of an external piece shrinks, whereas the second represents 
a degeneration of an internal sector. In both cases, the degenerate piece becomes a sphere
growing from the marked point.}}
\label{f:2pointd}
\end{center}
\end{figure}
Let us then study in this context the structure of the correlation functions we are going to compute.
The two-point function $d_n(\tau_1,\tau_2)$ of boundary operators with momenta $p_n$
is presented in fig.~\ref{f:2point}. One can see that $d_n$ is composed of the product
of $2n$ two-point functions $d_0$ of operators with vanishing boundary momentum.  
The precise relation is easy to write in the $\ell$-representation due to 
the factorization property of the path integral measure \cite{Kostov:1989eg,Kazakov:1991pt,Kostov:2002uq}
\be
\tilde d_n(\ell_1,\ell_2)=\int\limits_0^{\infty} d\ell'_1 \cdots d\ell'_{2n}\, 
\tilde d_0(\ell_1,\ell'_1) \,\tilde d_0(\ell'_1,\ell'_2)\,\cdots\, 
\tilde d_0(\ell'_{2n},\ell_2).
\label{reldnd}
\ee
From this representation one sees that singular contributions to $d_0$ give rise
to singularities of $d_n$, and the latter can have a much more complicated 
form. However, we know that the divergent contributions to $d_0$ arise from 
degenerate world sheets. In fact, since $d_n$ can be represented as a polynomial in $W(\tau)$, 
all its singularities can be interpreted in terms of 
degenerate configurations of some of the $2n+1$ disks composing the initial world sheet. 
The possible degenerate configurations are shown on fig.~\ref{f:2pointd},
and correspond either to a disk with two coinciding marked points or to a pinched disk,
together with a sphere growing from the marked point at the boundary.
These contributions have again the interpretation of contact terms and
should be neglected.  Our analysis suggests that, in order to eliminate all
of these contact terms, it is sufficient to remove the ones appearing in
the correlation functions with vanishing momenta corresponding
to each portion of the disk in fig.~\ref{f:2point}.
We will see in subsection~\ref{bndtwopsec} that this is indeed the case.

\begin{figure}
\begin{center}
\includegraphics[scale=0.8]{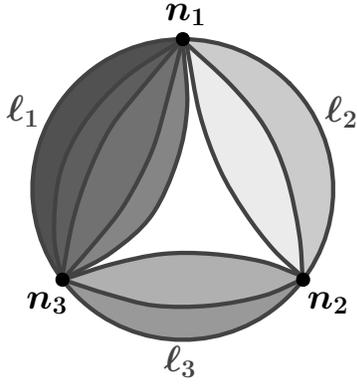}
\caption{{\small The three-point function $\tilde C_{n_1,n_2,n_3}(\ell_1,\ell_2,\ell_3)$.}}
\label{f:3point}
\end{center}
\end{figure}
\begin{figure}
\begin{center}
\includegraphics[scale=0.8]{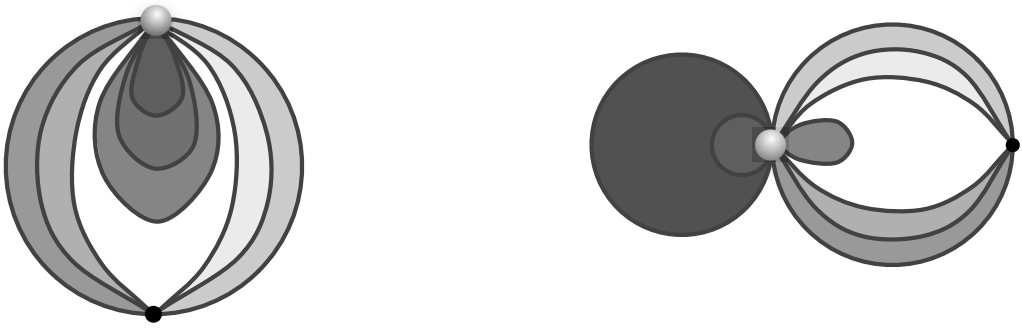}
\caption{{\small Possible degenerations of the three-point function.
The two cases correspond to shrinking of the boundary of an external 
or an internal sector, respectively. The degenerate piece appears as a sphere
attached to the point at the boundary where two of the three boundary operators collide.}}
\label{f:3pointd}
\end{center}
\end{figure}
A similar analysis can be applied to the case of the three-point function 
$C_{n_1,n_2,n_3}(\tau_1,\tau_2,\tau_3)$.
The representation in terms of non-intersecting lines is shown in fig.~\ref{f:3point}. 
We again see that the initial correlation function is reduced to a product
of two- and three-point correlators with vanishing momenta.
This is the origin of the representation of $C_{n_1,n_2,n_3}$
in terms of $W(\tau)$ that we will find below, and allows to interpret
each divergent contribution as a result of the degeneration of some of the disks
which divide the world sheet in fig.~\ref{f:3point}. The different possibilities are presented
in fig.~\ref{f:3pointd} and correspond to the same situations which we found in the case
of the two-point function. Notice that one could expect the appearance of  
the degenerate configuration with all three marked points coinciding,
which would give a world sheet in the form of a trefoil. However, one can show that the corresponding
divergent term in $C_{0,0,0}$ vanishes.

\begin{figure}
\begin{center}
\includegraphics[scale=0.8]{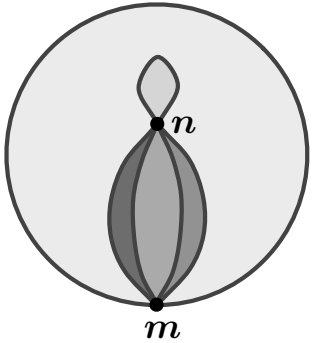}
\caption{{\small The bulk-boundary two-point function.}}
\label{f:bb}
\end{center}
\end{figure}

One can finally analyze the bulk-boundary correlator in the same way. Its
loop representation is depicted in fig.~\ref{f:bb}. In this case the lines 
corresponding to the boundary operator, instead of connecting two points of 
the boundary, end at the point where the bulk operator is inserted.
The naive counting would then tell that the correlation function should diverge as $\eps^{-n-m-1}$,
but we will see in subsection \ref{s:bulkboundary} that only the $1/\eps$ 
contribution is present. This can be explained from the fact, a consequence
of \eqref{Wtwobnd} and \eqref{expW}, that all terms divergent
as $\eps^{-k}$ with $k>1$ in the boundary two-point function are proportional to 
$\(\mub(\tau_1)-\mub(\tau_2)\)^{k-1}$ and, therefore, vanish at $\mub(\tau_1)=\mub(\tau_2)$.
Since the bulk operator cannot change the boundary conditions, 
the boundary cosmological constant indeed remains the same along the two boundaries of the 
disk formed by the outer lines coming from the two punctures. Thus, all contributions
with a higher degree of divergence than $1/\eps$ vanish.
The remaining divergent contribution turns 
out to be a polynomial 
in $\mub$, so it has the usual interpretation as contact term and has to be subtracted 
from the physical answer. 
 
After we remove all divergent contributions which can be interpreted as contact terms,
all correlation functions become finite. This is a non-trivial fact
which shows the consistency of the proposed renormalization. 
We interpret the resulting finite expressions as physical results
for the correlation functions in Liouville theory at $c_L=25$.
Notice that, in the cases where a comparison is possible, 
they also agree with the corresponding results of the discrete approach
obtained in \cite{Kostov:2002uq,Kostov:2003uh}. 

\subsection{Bulk one-point function}

First, we compute the simplest correlation function on the disk, namely the one-point bulk 
correlator~\eqref{onecorL}.\footnote{Some of the results of this subsection can be found in
\cite{Alexandrov:2004ks} where it is also shown that they agree with the results derived from 
matrix quantum mechanics.}
Substituting the expression for the Liouville momentum
from \eqref{alphbet} and replacing $s$ by $\tau$ using \eqref{taus}, one obtains
\be
U(\alpha_n|\tau)={1\over \pi b}\,[\pi\mul \gamma(b^2)]^{\hf\({1\over b^2}-1+{n\over bR}\)}
\textstyle\Gamma\(b^2-{nb \over R}\)\Gamma\(1-{1\over b^2}-{n\over bR}\)
\cosh\(\(1-{1\over b^2}-{n\over bR}\)\tau\).
\label{onecorLc}
\ee
Taking the limit $b\to 1$ and using the renormalization \eqref{tonep}
corresponding to the case of positive chirality, 
one finds, for $n\ge 1$ and generic compactification radius $R$,
\begin{align}
U_{c=1}(\alpha_n|\tau)&=
\under{\lim}{b\to 1}\[{1\over \pi^2 b}\,\muc^{\hf\({1\over b^2}-1+{n\over bR}\)}
\textstyle\Gamma\(1-b^2+{nb \over R}\)\Gamma\(1-{1\over b^2}-{n\over bR}\)
\cosh\(\(1-{1\over b^2}-{n\over bR}\)\tau\)\] 
\nonumber \\
&=-{\muc^{n\over 2R}\over {\pi n\over R}\,\sin{\pi n \over R}}\,
\cosh\({n \tau\over R}\).
\label{onecorLl}
\end{align}

The cases $R=1$ and $n=0$ should be treated separately. 
This can already be seen from the fact that the result \eqref{onecorLl} 
diverges at these values of the parameters.
Let us start from the case of the self-dual radius with $n\ne 0$.
The corresponding renormalized correlation function can be obtained from the first line in 
\eqref{onecorLl} substituting $R=b$. This gives
\be
\hat U(\alpha_n|\tau)= 
{1\over \pi^2 b}\,\muc^{\hf\({n+1\over b^2}-1\)}\Gamma\(n+1-b^2\)
\textstyle\Gamma\(1-{n+1\over b^2}\)
\cosh\(\(1-{n+1\over b^2}\)\tau\) .
\label{renonepb}
\ee
Substituting $b=1-\eps$ and expanding in $\eps$, one finds 
\be 
\hat U(\alpha_n|\tau)
\approx  {(-1)^{n+1}\over 2\pi^2 n}\,\muc^{{n\over 2}}
\[ \({ 1\over (n+1)\eps}+
\log\muc +c_U(n)\)\cosh(n \tau) +
2\tau \sinh(n \tau)
\] +O(\eps),
\label{oneapr}
\ee
where $c_U(n)$ is some $\tau$-independent function of $n$ which we are not interested in.
We observe that, despite the renormalization, the correlation function still diverges.
However, since $\cosh(n \tau)$ is a polynomial in $\cosh\tau$, 
the divergent term is a polynomial in the boundary cosmological constant $\mubc$. 
As we explained in the previous subsection, 
such terms are interpreted as contact terms and correspond to degenerate world sheets. 
We then ignore them and take the ``physical" one-point bulk
correlation function in the $c=1$ limit
to be just the last term in \eqref{oneapr}\footnote{Another possibility
would be to find some ``natural" subtraction defined before the limit $b=1$ is taken. 
For example, one can define the renormalized correlation function as follows
\begin{equation*}
\hat U^{\rm (ren)}(\alpha_n|\tau)=\hat U(\alpha_n|\tau)-
{1\over \pi^2 b}\,\Gamma(1+b^2-2b\alpha_n)\Gamma(2\alpha_n/b-1/b^2-1)
P^{(1)}_n\(\textstyle{\mubc\over \sqrt{\muc}} \),
\end{equation*}
where $P^{(1)}_n(\cos(x))=\cos(nx)$ is the Chebyshev polynomial of the first kind.
Then in the $c=1$ limit one obtains 
\begin{equation*}
{U^{(R=1)}_{c=1}(\alpha_n|\tau)} =
\mathop{\lim}\limits_{b\to1}
\hat U^{\rm (ren)}(\alpha_n|\tau)
={(-1)^{n+1}\over 2\pi^2 n}\,\muc^{{n\over 2}}
\( \log\muc \,\cosh(n\tau)+2\tau\sinh(n \tau)\) .
\end{equation*}
This result is finite but still contains a non-universal piece.
Therefore, we prefer to use the definition \eqref{physoneb}.}
\be
{U^{(R=1)}_{c=1}(\alpha_n|\tau)}=
{(-1)^{n+1}\over \pi^2 n}\,\muc^{{n\over 2}}\,\tau \,\sinh(n\tau).
\label{physoneb}
\ee

However, the case $n=0$ still remains singular.
To get the correlation function in this situation, we return to the representation 
\eqref{renonepb}, put $n=0$ and expand around $b=1$. The result reads
\be
{\hat U(\alpha_0|\tau)}\approx
-{1\over 4\pi^2 \eps^2}+{c_1-\log\muc\over 4\pi^2\eps}
 + \[ c_2+c_3\log\muc-{1\over 8\pi^2}\log^2\muc-{\tau^2\over 2\pi^2}\] +O(\eps),
\label{onecorcc}
\ee
$c_i$ being some unimportant constants.
The only term that depends on the boundary cosmological
constant in a non-trivial way is the last one in the finite part
of~\eqref{onecorcc}. Therefore, in this case the ``physical"
correlation function is
\be
{U_{c=1}(\alpha_0|\tau)}=-{\tau^2\over 2\pi^2}.
\label{physonebn}
\ee
Note that it does not coincide with the naive $n\to 0$ limit of \eqref{physoneb}, but
it agrees with the finite part of the $n\to 0$ limit of \eqref{onecorLl}.

Let us also give here the result for the one-point correlation function of the boundary 
cosmological constant operator. In this case we have to use the multiplicative
renormalization~\eqref{tonem} corresponding to negative chirality%
\footnote{The choice of chirality can be explained as follows.
$\hat W$ is obtained from the boundary three-point function
by integrating twice with respect to $\mubc$. The three-point function,
according to our convention, has two operators with positive and one with
negative chirality. Since each integral corresponds to removing one boundary cosmological
constant operator, which is an operator of positive chirality,
the resulting one-point function then necessarily corresponds to an operator of negative chirality.
This implies that, strictly speaking, $\hat W$ is not the one-point function of the
boundary cosmological constant operator, but rather of the operator
$\hat{\CB}_{-q}=e^{iqx}e^{b\phi}$. However, this reduces to $\hat{\CB}_{0}$ in the $c=1$ limit,
so we keep referring to $\hat{W}$ using this simpler terminology.

Another related subtlety is that in the theory with background charge the disk 
partition function is not well defined since the background momentum remains non-compensated.
},
and the correlator \eqref{oneW} becomes
\be
\hat W(\tau)=-\frac{2b}{\pi} \,\muc^{{1\over 2b^2}}\,
{\cosh(\tau /b^2)\over \sin(\pi /b^2)}.
\label{oneWren}
\ee
The substitution $b=1-\eps$ leads to the following expansion
\be
\hat W(\tau)\approx
{1\over \pi^2}
\({1\over  \eps}-{5\over 2}+\log\muc\)\mubc+
{2 \over \pi^2}\sqrt{\muc}\,\tau \sinh\tau +O(\eps)
\label{expW}
\ee
so that we can take 
\be
W_{c=1}(\tau)={2 \over \pi^2}\sqrt{\muc}\,\tau \sinh\tau.
\label{Wphys}
\ee
This is the well known expression for the one-point boundary function
of $c=1$ string theory which appears also as the resolvent in
the matrix model formulation.
Thus, we confirm the expectation that one should disregard
the contact terms in order to reproduce the physical results of the $c=1$
theory. Of course, the universal contribution \eqref{Wphys} 
agrees (up to contact terms) with \eqref{physonebn}.

\vspace{0.5em}

All correlation functions we are going to consider in the following
are simplified only if the restriction $R=1$ is imposed.
Therefore, from now on we will concentrate on this particular 
choice of the compactification radius and omit the label ``$(R=1)$".

\subsection{Bulk-boundary two-point function}
\label{s:bulkboundary}

Let us then proceed with the analysis of the bulk-boundary correlation function.
Substituting \eqref{alphbeto} into \eqref{bulkbnd}, one finds the following
expression for the bulk-boundary correlator
\begin{subequations}
\begin{align}
&\qquad\quad
R(\alpha_n,\beta_m|\tau)=  \left[\pi\mul \gamma(b^2)b^{2-2b^2}\right]^{{n+m+1\over 2 b^2}-1}
\,\CR_1 \, \CR_2,
\label{blkbndmom}\\
\CR_1&= {\Gamma_b^3\({m+1\over b}\)\Gamma_b\(b+{m-n\over b}\)\Gamma_b\({n+m+2\over b}-b\) \over
\Gamma_b\(b+{1\over b}\)\Gamma_b\({2m+1\over b}-b\)
\Gamma_b\(b-{m\over b}\)\Gamma_b\({n+1\over b}-b\)\Gamma_b\(2b-{n\over b}\)},
\label{bulkgam} \\
\CR_2&= -2\pi i\int\limits_{-i\infty}^{i\infty}dt\, e^{-2 t \tau/b}\
{\Sb_b\(t+b-{n+m+1\over 2b}\)\Sb_b\(t+{n-m+1\over 2b}\) \over
\Sb_b\(t+{n+m+3\over 2b}\)\Sb_b\(t+b+{m-n+1\over 2b}\)} .
\label{bulkbndb}
\end{align}
\end{subequations}
The second factor $\CR_1$ can be simplified by using the recursion relations
\eqref{recGb}. The result is
\be
\CR_1 = \frac{ (2\pi)^{2m}\, b^{n+m+1-2b^2-\frac{m+n+1}{b^2}}\, m!\,
\Gamma(b^2-n)\,\prod\limits_{k=m+1}^{2m} \Gamma\left(\frac{k}{b^2}\right)}
{\Gamma(2m+1-b^2)\prod\limits_{k=1}^{m} \Gamma\left(\frac{k}{b^2}\right)
\prod\limits_{k=-m}^{m} \Gamma\left(1+\frac{k}{b^2}\right)
\prod\limits_{k=-n}^{m-n-1} \Gamma\left(1+\frac{k}{b^2}\right)
\prod\limits_{k=n+1}^{m+n+1} \Gamma\left(\frac{k}{b^2}-1\right)}.
\label{Gratio}
\ee

The main non-trivial problem is to compute the integral $\CR_2$.
Thanks to the recursion relations among $\Sb_b$ functions,
the integrand can be simplified to just a product of sines:
\be
\CR_2= \frac{\pi i\,(-1)^{m}}{2^{2m+1}} \int\limits_{-i\infty}^{+i\infty} dt\
 \frac{e^{-{2 t \tau/b}}}
{\prod\limits_{k=\frac{n-m+1}{2}}^{\frac{n+m+1}{2}}
\[\sin\(\frac{\pi}{b}\left(t-\frac{k}{b}\right)\)
\sin\(\frac{\pi}{b}\left(t+\frac{k}{b}\right)\)\]}.
\label{Ibb}
\ee
The definition of the contour of integration in the $t$-plane
is the one explained in section \ref{subsec-cordisk}: the contour lies on the right of 
the poles of the integrand coming from the numerator, 
and to the left of the poles coming from the denominator. 
We also mentioned that this definition of the contour can become problematic 
if poles of the numerator become coincident with zeroes of the denominator. 
However, this never occurs if we limit ourselves to the case $n \ge m$, 
so in the following we assume this restriction. Closing the contour in the right half plane, 
the integral is then computed by picking residues at the following poles
\begin{equation}
	t = \frac{j}{b} + \ll b\,,\qquad \ll \ge 0
	\qquad\qquad \text{and} \qquad\qquad
	t = - \frac{j}{b} + \ll b\,,\qquad \ll > 0\,,
\end{equation}
where $j = \frac{n-m+1}{2},\ldots,\frac{m+n+1}{2}$. The computation of the residues gives:
\begin{equation}\label{Ibbresult}
\begin{split}
	\CR_2 &= \frac{(-1)^{m}\pi b}{2^{2m}}
		\sum\limits_{j=\frac{n-m+1}{2}}^{\frac{n+m+1}{2}}
		\frac{
		\sum\limits_{\ll=0}^{\infty} e^{-2\tau\left(\frac{j}{b^2}+\ll \right)}
		-\sum\limits_{\ll=1}^{\infty} e^{-2\tau\left(-\frac{j}{b^2}+\ll \right)}}		
		{\prod\limits_{k\neq j} \sin\(\frac{\pi}{b^2}\left(j-k\right)\)
		\prod\limits_{k} \sin\(\frac{\pi}{b^2}\left(j+k\right)\)}\\
		&= \frac{(-1)^{m}}{2^{2m}} \frac{\pi b}{1-e^{-2\tau}}
		\sum\limits_{j=\frac{n-m+1}{2}}^{\frac{n+m+1}{2}}
		\frac{ e^{-\frac{2j\tau}{b^2}} - e^{\frac{2j\tau}{b^2}-2\tau}}		
		{\prod\limits_{k\neq j} \sin\(\frac{\pi}{b^2}\left(j-k\right)\)
		\prod\limits_{k} \sin\(\frac{\pi}{b^2}\left(j+k\right)\)},
\end{split}
\end{equation}
where to avoid clutter of indices we have understood the range of the products over $k$, 
which is the same as in~\eqref{Ibb}.
Putting \eqref{Gratio} and~\eqref{Ibbresult} together, and 
taking into account the renormalization of the vertex operators \eqref{tone}, we finally get
\be\label{Rdegen}
\begin{split}
\hat R(\alpha_n,\beta_m|\tau) &=  
\frac{(-1)^m\, m!\, \pi^{2m-1}\, b^{-1}\,
\muc^{{n+m+1\over 2 b^2}-1} \,
\Gamma(1-b^2+n) \Gamma\(\frac{2m+1}{b^2}-1\)
\prod\limits_{k=m+1}^{2m} \Gamma\left(\frac{k}{b^2}\right)}
{\Gamma(2m+1-b^2) \prod\limits_{k=1}^{m} \Gamma\left(\frac{k}{b^2}\right)
\prod\limits_{k=-m }^{m} \Gamma\left(1+\frac{k}{b^2}\right)
\prod\limits_{k=-n}^{m-n-1} \Gamma\left(1+\frac{k}{b^2}\right)
\prod\limits_{k=n+1}^{m+n+1} \Gamma\left(\frac{k}{b^2}-1\right)} 
\\
& \quad \times  \frac{1}{1-e^{-2\tau}}
\sum\limits_{j=\frac{n-m+1}{2}}^{\frac{n+m+1}{2}}
\frac{ e^{ - \frac{2j\tau}{b^2}} - e^{\frac{2j\tau}{b^2}-2\tau}}		
{\,\,\sin{\frac{2\pi j}{b^2}}\prod\limits_{k=\frac{n-m+1}{2} \atop k\neq j}^{\frac{n+m+1}{2}} 
\[\sin\(\frac{\pi}{b^2}\left(j-k\right)\)\,
\sin\(\frac{\pi}{b^2}\left(j+k\right)\)\]} .
\end{split}
\ee

Let us now extract the leading terms in the $b\to 1$ limit. We find
\begin{align}\label{renblukbnd}
\hat R&(\alpha_n,\beta_m|\tau)\\
&\approx {(-1)^{n+1}\,m!\, \muc^{{n+m-1\over 2 }}\over 2\pi^2 \eps}\,
\(\prod\limits_{k=1}^{m} {\Gamma\left(k+m\right)\Gamma\left(k+n-m+1\right)
\over \Gamma\left(k\right)\Gamma\left(k+n\right)}\)
\, \bigl(1+ c_R(n,m)\eps\bigr) \CR_{n,m}(\tau) +O(\eps),\nonumber
\end{align}
where
\beq 
&&\hspace{-0.6cm}
\CR_{n,m}(\tau)=  \sum\limits_{j=0}^{m}{v_j(n,m)}\,
\frac{\sinh\(\(1 -\frac{2j+n-m+1}{b^2}\)\tau\)}		
{\sinh\tau} 
\label{lastl}
\\ 
& & \hspace{-0.4cm}\approx
-\sum\limits_{j=0}^{m}v_j(n,m)\[
\frac{\sinh\(\(2j+n-m\)\tau\)}		
{\,\sinh\tau}
+2(2j+n-m+1)\eps\,
\frac{\tau\cosh\( \(2j+n-m\)\tau\)}{\sinh\tau} \] +O(\eps)
\nonumber 
\eeq
with $v_j(n,m)= {(-1)^{j}\,(j+n-m)!\over j!\,(m-j)!\,(j+n+1)!}$. For simplicity
we also included the term proportional to $\log\muc$ into the constant $c_R(n,m)$.
As a result, the leading term of \eqref{renblukbnd} is proportional to 
\be
\frac{1}{\eps} \sum\limits_{j=0}^m
{v_j(n,m)}\, P^{(2)}_{2j+n-m-1} \(\textstyle{\mubc\over \sqrt{\muc}} \),
\label{leadprop}
\ee
where $P^{(2)}_{n-1}(\cos(x))={\sin(nx)\over sin(x)}$ 
is the Chebyshev polynomial of the second kind.
Thus, the leading term is divergent but is a polynomial in $\mubc$:
it can then be interpreted as a contact term and disregarded.
The physical contribution can be read off the second term in \eqref{lastl}
and is given by
\begin{equation}
\begin{split}
\label{nnblkbnd}
R_{c=1}(\alpha_n,\beta_m|\tau)&=
{m!\over \pi^2}\, \muc^{{n+m-1\over 2 }} 
\(\prod\limits_{k=1}^{m} {\Gamma\left(k+m\right)\Gamma\left(k+n-m+1\right)
\over \Gamma\left(k\right)\Gamma\left(k+n\right)}\)\\ 
&\times
\sum\limits_{j=0}^{m}
\frac{(-1)^{n+j}\,(2j+n-m+1)(j+n-m)!}		
{(m-j)!\, (n+j+1)!}\,\frac{\tau \cosh\((2j+n-m)\tau \)}{ \sinh\tau}.
\end{split}
\end{equation}
In the case where the condition on the momenta 
following from the matter sector is taken into account, requiring $n=m$,
this expression gets some additional simplifications, becoming
\be
R_{c=1}(\alpha_n,\beta_n|\tau)= 
{\muc^{n-{1\over 2 }} \over \pi^2}
\sum\limits_{j=0}^{n}
\frac{(-1)^{n+j}\,(2j+1)\,(n!)^2}		
{(n-j)!\, (n+j+1)!}\,\frac{\tau \cosh\(2j\tau \)}{ \sinh\tau}.
\label{nnblkbndsim}
\ee

Let us also perform a consistency check, namely that the result \eqref{renblukbnd}
together with \eqref{lastl} reproduces the bulk one-point function found in 
\eqref{oneapr} for $n>0$ and in \eqref{onecorcc} for $n=0$.
For $m=0$ one finds
\be
\hat R(\alpha_n,\beta_0|\tau)\approx{(-1)^n\muc^{{n-1\over 2 }}\over 2\pi^2 }
\[\frac{\sinh\(n\tau \)}		
{(n+1)\eps\,\sinh\tau}\(1+c_R(n,0)\eps\)
+\frac{2\tau\cosh\(n\tau \)}		
{\sinh\tau} \] .
\label{blkbndmzero}
\ee
Then the integral with respect to $-\mubc$ is easily calculated, yielding
\be
-\int d\mubc \, \hat R(\alpha_n,\beta_0|\tau)\approx
\left\{\begin{array}{ll}
{(-1)^{n+1}\muc^{{n\over 2 }}\over 2\pi^2 n}
\[\({1\over (n+1)\eps}+ c_R(n)\)\cosh\(n\tau\)
+2 \tau\sinh\(n\tau \)\] & n>0,
\\
-{\tau^2\over 2\pi^2}
& n=0,
\end{array}
\right.
\label{checkone}
\ee
in perfect agreement with \eqref{oneapr} and with the universal part 
of \eqref{onecorcc}. The non-universal part for $n=0$
does not depend on $\tau$ and cannot be fixed starting from the integral \eqref{checkone}.

\subsection{Boundary two-point function}
\label{bndtwopsec}

The next function which we are going to consider is the boundary two-point correlator.
Although it follows from the three-point function, it is useful to start with this
simpler quantity since it allows to illustrate the main non-trivial points
of the renormalization leading to a finite answer in the limit $b\to 1$. 

The explicit expression for the two-point function was found in appendix \ref{C}
and is given in \eqref{twofunrel} together with \eqref{twopbnd}.
Using~\eqref{alphbeto} and replacing $s_i$ by $\tau_i$, one obtains
\be
d(\beta_n|\tau_1,\tau_2) =
{2\({2n+1\over b}-b\)
\left[\pi \mu \gamma(b^2) \,b^{2-2b^2}\right]^{\hf\({2n+1\over b^2}-1\)}
\Gamma_b\(b-{2n+1\over b}\)\Gamma_b^{-1}\({2n+1\over b}-b\)
\over \Sb_b\(b-{n\over b}+i{\tau_1+\tau_2\over 2\pi b}\)
\Sb_b\(b-{n\over b}-i{\tau_1+\tau_2\over 2\pi b}\)
\Sb_b\(b-{n\over b}+i{\tau_1-\tau_2\over 2\pi b}\)
\Sb_b\(b-{n\over b}-i{\tau_1-\tau_2\over 2\pi b}\) }.
\label{twobbbb}
\ee
By using the properties of the special functions,
the factor in the numerator can be transformed as follows
\be
b^{(1-b^2)\({2n+1\over b^2}-1\)}
{\Gamma_b\(b-{2n+1\over b}\)\over \Gamma_b\({2n+1\over b}-b\)}=
{2^{-2n}\pi \over \Gamma\({2n+1}-b^2\) \Gamma\(2n+1\over b^2\)}
\prod\limits_{k=1}^{2n+1}{1\over \sin{\pi k\over b^2}} .
\ee
For the denominator in \eqref{twobbbb}, we get
\begin{equation}
\frac{\Sb_b \left(\frac{n+1}{b}+i\frac{\tau_1+\tau_2}{2\pi b}\right)
\Sb_b \left(\frac{n+1}{b}+i\frac{\tau_1-\tau_2}{2\pi b}\right)}
{\Sb_b \left(b-\frac{n}{b}+i\frac{\tau_1+\tau_2}{2\pi b}\right)
\Sb_b \left(b-\frac{n}{b}+i\frac{\tau_1-\tau_2}{2\pi b}\right)}
= \frac{2^{4n} \prod\limits_{k=-n}^{n}\[
\sinh\(\frac{\tau_1+\tau_2+2\pi i k}{2b^2}\)
\sinh\(\frac{\tau_1-\tau_2+2\pi i k}{2b^2}\)\]}
{\sinh\left(\frac{\tau_1+\tau_2}{2}\right)
\sinh\left(\frac{\tau_1-\tau_2}{2}\right)}\,.
\end{equation}
Then the final expression, without special functions and renormalized accordingly to \eqref{tone}, 
turns out to be
\begin{equation}
\hat d(\beta_n|\tau_1,\tau_2) =  \frac{2b}{\pi}\,
\muc^{\frac{1}{2}\left(\frac{2n+1}{b^2}-1\right)}\, 
\frac{ \prod\limits_{k=1}^{2n+1}{1\over \sin{\pi k\over b^2}} \,
\prod\limits_{k=-n}^{n}\(
\cosh\(\tau_1 +2\pi i k \over b^2\)-\cosh{\tau_2\over b^2}\) }
{ \cosh\tau_1-\cosh\tau_2 }.
\end{equation}

Remarkably, this expression can be rewritten in terms of the (renormalized)
one-point functions $\hat W_i \equiv \hat W(\tau_i)$, $i=1,2$, 
given in~\eqref{oneWren}.
The result reads
\be
\hat d(\beta_n|\tau_1,\tau_2) =  - a_n \, 
\frac{\hat W_1  -\hat W_2}{ \muc_{12} }\,
\prod\limits_{k=1}^{n}\( \hat W^2_1+\hat W^2_2
-2\cos{2k\pi\over b^2}\, \hat W_1\hat W_2
-\frac{4b^2}{\pi^2}\muc^{1\over b^2}\,{\sin^2{2\pi k\over b^2}\over \sin^2{\pi\over b^2}}\),
\label{Wtwobnd}
\ee
where we used the definition \eqref{notmu} and introduced
\be
a_n= \left(\frac{\pi}{2b}\right)^{2n}\,
\prod\limits_{k=1}^{2n}{\sin{\pi\over b^2}\over \sin{\pi (k+1)\over b^2}}.
\label{const_c}
\ee
In particular, for $n=0$, namely for the two-point function of $\CB_0^{\tau_1\tau_2}$ and $\CB_{-q}^{\tau_2\tau_1}$, we obtain%
\footnote{In the limit $\tau_1=\tau_2$, \eqref{twocos} reproduces the result \eqref{twobndsim}
for the correlator of two cosmological constant operators
multiplied by the renormalization factor ${1\over \pi^2}\Gamma(1-b^2)\Gamma\left(1/b^2-1\right)$.}
\be
\hat d(b|\tau_1,\tau_2)=-\frac{\hat W_1-\hat W_2}{\muc_{12}}.
\label{twocos}
\ee
Note that the expressions \eqref{Wtwobnd} and \eqref{twocos}, 
with a slightly different normalization, appeared in \cite{Kostov:2003uh}
where they were derived from the discrete approach
based on the SOS model and the loop gas representation.

To extract the physical result for the $c=1$ theory, let us note that all coefficients
in \eqref{Wtwobnd} in this limit remain finite. In particular
\be
\lim\limits_{b\to 1} a_n= {(-1)^{n}\pi^{2n}\over 2^{2n}(2n+1)!}.
\label{const_clim}
\ee
Thus, to get a finite result, it is enough to renormalize the one-point functions 
$\hat W(\tau)$ appearing in \eqref{Wtwobnd}. The interpretation of such a
renormalization was given in section \ref{secuniv}, where it was shown that it corresponds 
to neglecting world sheet configurations with two marked points on the boundary
close to each other so that the world sheet degenerates. Replacing then $\hat W(\tau)$
by its physical part in the $c=1$ limit given in \eqref{Wphys}, one finds
\beq
&& d_{c=1}(\beta_n|\tau_1,\tau_2) = 
{(-1)^{n+1}\pi^{2n}\over 2^{2n}(2n+1)!}
\frac{W_{c=1}(\tau_1)-W_{c=1}(\tau_2)}
{ \mubc(\tau_1)-\mubc(\tau_2) }
\prod\limits_{k=1}^{n}\( \(W_{c=1}(\tau_1)-W_{c=1}(\tau_2)\)^2
-{16k^2\muc\over \pi^2} \)
\nonumber \\
& & \qquad\quad 
={2(-1)^{n+1}\muc^{2n}\over \pi^{2n+2}(2n+1)!}
{\tau_1\sinh\tau_1-\tau_2\sinh\tau_2 \over \cosh\tau_1-\cosh\tau_2}
\prod\limits_{k=1}^{n}\( \(\tau_1\sinh\tau_1-\tau_2\sinh\tau_2\)^2
-(2\pi k)^2 \).
\label{dphys}
\eeq
It is clear that the divergent contribution to $\hat W(\tau)$ gives rise
to a series of divergent contributions to $\hat d(\beta_n|\tau_1,\tau_2)$. 
Only the leading contribution 
is an entire function of $\mubc$, whereas all others have a non-trivial functional dependence
on $\tau_i$. Nevertheless, all of them correspond to contact terms according
to our reasoning in section \ref{secuniv}. The only physical 
contribution is given by the finite expression \eqref{dphys}. 

\subsection{Boundary three-point function}
\label{bndthreesec}

Now we turn to the most complicated correlation function, the three-point correlator \eqref{threepbnd}. 
Substituting the momenta \eqref{alphbeto} and replacing $s_i$ by $\tau_i$, one obtains
\begin{subequations}
\begin{align}
	&\qquad\qquad C(\beta_{n_1},\beta_{n_2},\beta_{n_3}|\tau_1,\tau_2,\tau_3) =
		\left[\pi\mul \gamma(b^2)b^{2-2b^2}\right]^{{1+\sum n_i\over 2b^2}-1}\,
		\mathcal{C}_1\, \mathcal{C}_2\, \mathcal{C}_3,
\label{threepbndb}\\
\mathcal{C}_1 &= \frac{\Gamma_b \left( b + \frac{n_1-n_2-n_3}{b} \right)
		\Gamma_b \left( - b + \frac{n_1+n_2+n_3+2}{b} \right)
		\Gamma_b \left( \frac{n_1+n_2-n_3+1}{b} \right)
		\Gamma_b \left( \frac{n_1-n_2+n_3+1}{b} \right)}
		{\Gamma_b \left( b + \frac{1}{b} \right)
		\Gamma_b \left( - b + \frac{2n_1+1}{b} \right)
		\Gamma_b \left( - b + \frac{2n_2+1}{b} \right)
		\Gamma_b \left( - b + \frac{2n_3+1}{b} \right)}\,, \label{C1}\\
\mathcal{C}_2 &=
		{\Sb_b\({n_3+1\over b}+i{\tau_1-\tau_3\over 2\pi b}\)
		\Sb_b\({n_3+1\over b}-i{\tau_1+\tau_3\over 2\pi b}\)
		\over \Sb_b\(b-{n_2\over b}+i{\tau_2-\tau_3\over 2\pi b}\)
		\Sb_b\(b-{n_2\over b}-i{\tau_2+\tau_3\over 2\pi b}\)}\,,\label{C2}\\
\mathcal{C}_3 &=
        - 4\pi i
		\int\limits_{-i\infty}^{i\infty}dt\, \prod\limits_{i=1}^4 
		\frac{\Sb_b(t+U_i)}{\Sb_b(t+V_i)},
\label{C3}
\end{align}
\end{subequations}
and
\bea{rclcrcl}{parametb}
U_1&=&{n_1+1\over b}+i{\tau_1+\tau_2\over 2\pi b}, & \qquad &
V_1&=&{n_1+n_3+2\over b}+i{\tau_2-\tau_3\over 2\pi b},
\\
U_2&=&{n_1+1\over b}-i{\tau_1-\tau_2\over 2\pi b}, & \qquad &
V_2&=&{n_1-n_3+1\over b}+b+i{\tau_2-\tau_3\over 2\pi b},
\\
U_3&=&b-{n_2\over b}+i{\tau_2-\tau_3\over 2\pi b}, & \qquad &
V_3&=&b+{1\over b}+{i\tau_2\over \pi b},
\\
U_4&=&{n_2+1\over b}+i{\tau_2-\tau_3\over 2\pi b}, & \qquad &
V_4&=& b+{1\over b}.
\\
\eea

Let us start by simplifying $\mathcal{C}_1$. In the following we will restrict 
ourselves to the case when the momenta 
$n_i$ satisfy the triangle inequality, $|n_2- n_3|\le n_1 \le n_2+ n_3$.
This allows to avoid considering many different cases which can not be treated 
simultaneously and is consistent with the momentum conservation imposed by the matter sector. 
One then obtains the following result for the quantity in~\eqref{C1}:
\begin{equation}
\mathcal{C}_1 = \frac{
b^{3+3n_2+3n_3-n_1-2b^2-{\sum n_i+1\over b^2}}}
{(2\pi)^{n_2+n_3-n_1-1}}
\frac{\prod\limits_{k=0}^{n_2+n_3-n_1} \Gamma \left( 1- \frac{k}{b^2} \right)\ 
\prod\limits_{k=n_1+n_2-n_3+1}^{2n_2} \Gamma \left( \frac{k}{b^2} \right)\ 
\prod\limits_{k=n_1+n_3-n_2+1}^{2n_3} \Gamma \left( \frac{k}{b^2} \right)}
{\Gamma(2n_2+1-b^2)\Gamma(2n_3+1-b^2)
\prod\limits_{k=2n_1+1}^{n_1+n_2+n_3+1} \Gamma \left( -1+ \frac{k}{b^2} \right)}.
\label{Conesim}
\end{equation}

Now we pass to the evaluation of $\mathcal{C}_3$. As a first step, 
we can eliminate four of the eight $\Sb_b$ functions appearing in the integral, obtaining
\begin{equation}\label{C32}
\begin{split}
\mathcal{C}_3 =&\,
\frac{\pi i(-1)^{\sum n_i}}{4^{n_1}} \int\limits_{-i\infty}^{i\infty}dt\, 
\frac{\Sb_b\(t+{n_1+1\over b}+i{\tau_1+\tau_2\over 2\pi b}\)\, 
\Sb_b\(t+{n_1+1\over b}-i{\tau_1-\tau_2\over 2\pi b}\)}
{\Sb_b\(t+{n_2+1\over b}+i{\tau_2-\tau_3\over 2\pi b}\)\, 
\Sb_b\(t+b+{1\over b}\)}
\\
&\times 
\textstyle
\[\prod\limits_{k=n_2+1}^{n_1+n_3+1} \sin\left(\frac{\pi}{b}\left(t+i\frac{\tau_2-\tau_3}{2\pi b}
+\frac{k}{b}\right)\right)
\prod\limits_{k=-n_2}^{n_1-n_3} \sin\left(\frac{\pi}{b}\left(t+i\frac{\tau_2-\tau_3}{2\pi b}
+\frac{k}{b}\right)\right)\]^{-1} .
\end{split}
\end{equation}
The contour of integration in the complex $t$-plane is defined as usual 
to lie on the right (left) of the poles 
of the integrand coming from poles (zeros) of the $\Sb_b$ functions in the numerator 
(denominator) of~\eqref{C3}. Due to our condition on the momenta, 
the poles coming from the numerator and from the denominator 
never come to collide, and the contour is well defined. 

However, there is an important difference between this integral and the one 
appearing in the expression for the bulk-boundary correlator.
Whereas in the latter case we could close the integration contour in the right half plane
due to the exponential fall off of the integrand, this cannot be done in the present case.
In a similar situation, the one of the derivation of the boundary two-point function from the three-point
function in appendix \ref{C}, as a solution to this problem we proposed to use some ``shift equations" 
characterizing the result of integration. 
We apply the same strategy in appendix \ref{secshifts} for the case at hand. We demonstrate that
the equations we derive for the integral \eqref{C32} are satisfied by the expression that one obtains by closing the contour of integration in the right half 
plane and taking into account only contributions from the zeros of the sine functions
in \eqref{C32}. Notice that, according to these  ``effective integration rules'',
the poles of the $\Sb_b$-functions, which
have a much more complicated form, remarkably do not contribute to the result.

In summary, in order to compute the integral 
we have to pick residues at the following two series of single poles 
\be
t_{j\ll} = - i\frac{\tau_2-\tau_3}{2\pi b} - \frac{j}{b} + \ll b, \qquad
\left\{ 
\begin{array}{ll}
j = n_2+1,\ldots,n_1+n_3+1 ,& \ll > 0, 
\\
j = -n_2,\ldots,n_1-n_3,& \ll \ge 0.
\end{array}
\right.
\label{3ptres}
\ee
For both series the residues of the integrand $\mathcal{I}(t)$  
can be written as
\be
\res\limits_{t=t_{j\ll}}\mathcal{I}(t)
=\frac{(-1)^{\sum n_i}\, i b}
{4^{n_1} \prod\limits_{ k\neq j} 
\sin\(\frac{\pi}{b^2}\(k-j\)\)}
\times \mathcal{S}_{j\ll},
\label{res1}
\ee
where $k$ in the product runs through the union of two intervals
$k\in[-n_2,n_1-n_3]\cup[n_2+1,n_1+n_3+1] $ and
the quantity $\mathcal{S}_{j\ll}$ is defined as
\begin{equation}\label{Sjl}
	\mathcal{S}_{j\ll} =
		\frac{\Sb_b\left(i\frac{\tau_1+\tau_3}{2\pi b}+\frac{n_1+1-j}{b}+\ll b\right)\ 
		\Sb_b\left(-i\frac{\tau_1-\tau_3}{2\pi b}+\frac{n_1+1-j}{b}+\ll b\right)}
		{\Sb_b\left(i\frac{\tau_2+\tau_3}{2\pi b}+\frac{1-j}{b}+(\ll+1) b\right)
		\Sb_b\left(-i\frac{\tau_2-\tau_3}{2\pi b}+\frac{1-j}{b}+(\ll+1) b\right)}\,.
\end{equation}
All of the special functions in $\mathcal{S}_{j\ll}$ can be eliminated by 
multiplying~\eqref{Sjl} by~\eqref{C2}, getting
\be
\mathcal{C}_2\, \mathcal{S}_{j\ll}=
{4^{\sum n_i}}\pi \, \mathcal{A}_{j}\mathcal{B}_{\ll},
\label{defAB}
\ee
where
\begin{align}
\mathcal{A}_{j} &= \textstyle\prod\limits_{k=-n_3}^{n_1-j} \left[
		\sinh\(\frac{\tau_1+\tau_3-2\pi i k }{2b^2} \)
		\sinh\(\frac{\tau_1-\tau_3+2\pi i k }{2b^2} \)\right]
 \prod\limits_{k=1-j}^{n_2} \left[
		\sinh\(\frac{\tau_2+\tau_3-2\pi i k }{2b^2} \)
		\sinh\(\frac{\tau_2-\tau_3+2\pi i k }{2b^2} \)\right],
\label{defAA}\\
\mathcal{B}_{\ll}&= \frac{\prod\limits_{k=0}^{\ll-1} \left[
		\sinh\left( \frac{\tau_1+\tau_3}{2} - \pi i k b^2\right)
		\sinh\left( \frac{\tau_1-\tau_3}{2} + \pi i k b^2\right)\right]}
		{\sinh\left( \frac{\tau_1+\tau_3}{2}\right)
		\sinh\left( \frac{\tau_1-\tau_3}{2}\right)
		\prod\limits_{k=0}^{\ll} \left[
		\sinh\left( \frac{\tau_2+\tau_3}{2} - \pi i k b^2\right)
		\sinh\left( \frac{\tau_2-\tau_3}{2} + \pi i k b^2\right)\right]}.
\label{defBB}
\end{align}
Collecting all factors from~\eqref{res1}-\eqref{defBB}, we can write 
the final answer for the product $\mathcal{C}_2\, \mathcal{C}_3$:
\begin{equation}\label{3ptfinal}
\begin{split}
\mathcal{C}_2\, \mathcal{C}_3 &= 2\pi b\, (-1)^{\sum n_i} 4^{n_2+n_3-1}
\Bigg\{\sum_{\ll=1}^{\infty}\mathcal{B}_{\ll} 
\sum_{j=n_2+1}^{n_1+n_3+1} 
\frac{\mathcal{A}_{j}}
{\prod\limits_{k=n_2+1 \atop k\neq j}^{n_1+n_3+1} \sin\(\frac{\pi}{b^2}\(k-j\)\)
\prod\limits_{k=-n_2}^{n_1-n_3} \sin\(\frac{\pi}{b^2}\(k-j\)\)}
\\
&\qquad\qquad\qquad\qquad\qquad + \sum_{\ll=0}^{\infty}\mathcal{B}_{\ll}
\sum_{j=-n_2}^{n_1-n_3} 
\frac{\mathcal{A}_{j}}
{\prod\limits_{k=n_2+1 }^{n_1+n_3+1} \sin\(\frac{\pi}{b^2}\(k-j\)\)
\prod\limits_{k=-n_2 \atop k\neq j}^{n_1-n_3} \sin\(\frac{\pi}{b^2}\(k-j\)\)}\Bigg\}.
\end{split}
\end{equation}

It turns out that the sums over $\ll$ in \eqref{3ptfinal} can be calculated explicitly.
For this let us introduce the shift operator in $\tau_3$
\be
\Delta_3 =e^{-2\pi i b^2 \p_{\tau_3}}.
\label{shift}
\ee
Using this operator, one can write
\be
\sum_{\ll=1}^{\infty}\mathcal{B}_{\ll}  =
{4\sqrt{\muc}\over \hat\mu_{23}}\,
{\sum\limits_{\ll=0}^{\infty} 
\( \Delta_3 \,{\cosh\tau_1-\cosh\tau_3\over \cosh\tau_2-\cosh( \tau_3-2\pi ib^2)}\)^l
\over 
\cosh\tau_2-\cosh(\tau_3-2\pi ib^2)}
=
{4 \sqrt{\muc} \over \hat\mu_{23}}\,
{\[ 1- \Delta_3 \,{\cosh\tau_1-\cosh\tau_3\over 
\cosh\tau_2-\cosh(\tau_3-2\pi ib^2)}\]^{-1}
\over 
\cosh\tau_2-\cosh(\tau_3-2\pi ib^2)}.
\label{sumBl}
\ee
Then from the following identity 
\be
\[ 1- \Delta_3 \,{\cosh\tau_1-\cosh\tau_3\over 
\cosh\tau_2-\cosh(\tau_3-2\pi ib^2)}\] {\cosh\tau_2-\cosh(\tau_3-2\pi ib^2)\over 
\cosh\tau_1-\cosh\tau_2}=-1
\label{identBl}
\ee
it follows that 
\begin{equation}
\sum_{\ll=1}^{\infty} {\mathcal{B}}_\ll
= -\frac{4\muc}{\hat\mu_{12}\,\hat\mu_{23}},
\qquad
\sum_{\ll=0}^{\infty} {\mathcal{B}}_\ll
=  -\frac{4\muc}{\hat\mu_{12}\,\hat\mu_{13}}.
\label{infsum}
\end{equation}
Remarkably, the results do not depend on $b$.
This means that they will not generate an expansion in $\eps=1-b$.

Note that the quantities $\CA_j$ defined in \eqref{defAA} can be written in terms of
the renormalized one-point functions $\hat W(\tau)$ given in~\eqref{oneWren}:
\be
\CA_j=\[-{4 b\muc^{1\over 2b^2}\over  \pi\sin(\pi/b^2)}\]^{-\sum n_i-1}
\textstyle\prod\limits_{k=-n_3}^{n_1-j} \left(
\hat W(\tau_1)-\hat W(\tau_3-2\pi i k)\right)
 \prod\limits_{k=1-j}^{n_2}  \left(
\hat W(\tau_2)-\hat W(\tau_3-2\pi i k)\right).
\label{WAA}
\ee
Thus, after the multiplicative renormalization \eqref{tone}, where we choose
the third operator to be the one with negative chirality,
the expression for the full three-point function reads as follows
\begin{equation}\label{3ptfullfinal}
\begin{split}
& \hat C(\beta_{n_1},\beta_{n_2},\beta_{n_3}|\tau_1,\tau_2,\tau_3) 
\\
& \quad
=\,
\CC_0\,
\Bigg\{\frac{1}{\hat\mu_{12}\,\hat\mu_{23}} 
\sum_{j=n_2+1}^{n_1+n_3+1} 
\frac{\prod\limits_{k=-n_3}^{n_1-j} \left(
\hat W(\tau_1)-\hat W(\tau_3-2\pi i k)\right)
 \prod\limits_{k=1-j}^{n_2}  \left(
\hat W(\tau_2)-\hat W(\tau_3-2\pi i k)\right)}
{\prod\limits_{k=n_2+1 \atop k\neq j}^{n_1+n_3+1} \sin\(\frac{\pi}{b^2}\(k-j\)\)
\prod\limits_{k=-n_2}^{n_1-n_3} \sin\(\frac{\pi}{b^2}\(k-j\)\)}
\\
&\qquad\quad + \frac{1}{\hat\mu_{12}\,\hat\mu_{13}}
\sum_{j=-n_2}^{n_1-n_3} 
\frac{\prod\limits_{k=-n_3}^{n_1-j} \left(
\hat W(\tau_1)-\hat W(\tau_3-2\pi i k)\right)
 \prod\limits_{k=1-j}^{n_2}  \left(
\hat W(\tau_2)-\hat W(\tau_3-2\pi i k)\right)}
{\prod\limits_{k=n_2+1 }^{n_1+n_3+1} \sin\(\frac{\pi}{b^2}\(k-j\)\)
\prod\limits_{k=-n_2 \atop k\neq j}^{n_1-n_3} \sin\(\frac{\pi}{b^2}\(k-j\)\)}\Bigg\},
\end{split}
\end{equation}
where the overall coefficient is 
\begin{align}
\CC_0&= \frac{\Gamma(2n_1+1-b^2)\,
\Gamma(2n_2+1-b^2)\, \Gamma\left( \frac{2n_3+1}{b^2}-1 \right)}
{2^{2n_1+1}\, \pi^{1-\sum n_i}\, b^{3+2\sum n_i-2b^2-{1+\sum n_i\over b^2}}}
\(\sin\frac{\pi}{b^2}\)^{\sum n_i+1} \mathcal{C}_1
\label{overcoef3}
\\
&= 
\frac{\Gamma(2n_1+1-b^2) \Gamma\left( \frac{2n_3+1}{b^2}-1 \right)
\prod\limits_{k=1}^{n_2+n_3-n_1} \Gamma \left( 1- \frac{k}{b^2} \right)\ 
\prod\limits_{k=n_1+n_2-n_3+1}^{2n_2} \Gamma \left( \frac{k}{b^2} \right)\ 
\prod\limits_{k=n_1+n_3-n_2+1}^{2n_3} \Gamma \left( \frac{k}{b^2} \right)}
{2^{\sum n_i}\, \pi^{-2n_1}\, b^{3n_1-n_2-n_3}\, 
\(\sin\frac{\pi}{b^2}\)^{-\sum n_i-1}\, \Gamma(2n_3+1-b^2)
\prod\limits_{k=2n_1+1}^{n_1+n_2+n_3+1} \Gamma \left( -1+ \frac{k}{b^2} \right)}.
\nonumber
\end{align}
Although this result is still lacking of the explicit cyclic symmetry under
simultaneous permutations of $n_i$ and $\tau_i$, we checked
on examples for low values of momenta that the cyclic symmetry is indeed present.
Several explicit expressions can be found in appendix \ref{explicthree}. 
From these examples one also observes that all imaginary terms produced by shifts of
the argument of $\hat W$ in \eqref{3ptfullfinal} are canceled and the final
expressions can be written solely in terms of $\hat W(\tau_i)$.
Actually, this is a consequence of the cyclic symmetry, since the imaginary terms are odd functions
of $\tau_3$, whereas the function \eqref{3ptfullfinal} is even in $\tau_1$ and $\tau_2$,
which means that it should also be even in $\tau_3$.
This is a crucial result for the $c=1$ limit. Unfortunately, we cannot
prove it in general explicitly due to the complicated combinatorics 
involved in \eqref{3ptfullfinal}, but since it follows from the fundamental symmetry 
of the three-point function, it ought to be true. 

The general structure of the three-point function evaluated for degenerate momenta,
namely being a polynomial in $\hat W(\tau_i)$ divided by the differences of the boundary
cosmological constants, was derived from the microscopic approach in \cite{Kostov:2003uh}.
However, the explicit expression for the polynomial was missing.
Although our result is also not as explicit as one might wish, in order to
obtain this polynomial it is enough to expand $\hat W(\tau_3-2\pi i k)$, cancel
all imaginary terms and rewrite the result in terms of $\hat W(\tau_3)$, as we did
for some cases in appendix \ref{explicthree}.

The fact that the three-point function is a polynomial in $\hat W(\tau_i)$
allows to apply the renormalization procedure described in section \ref{secuniv}
and already successfully used for the boundary two-point function.
Let us first study the limit $b\to 1$ of various coefficients appearing in \eqref{3ptfullfinal}.
Expanding $b=1-\eps$, the leading part of the factor $\mathcal{C}_0$ given in \eqref{overcoef3}
reads
\be
\mathcal{C}_0\approx
(2\pi\eps)^{2n_1+1}\,
(-1)^{\hf(n_2+n_3-n_1)(n_2+n_3-n_1-1)+1}\,\(\pi \over 2\)^{\sum n_i}\,
\frac{\prod\limits_{k=n_1+n_2-n_3}^{2n_2-1} k!\ \prod\limits_{k=n_1+n_3-n_2}^{2n_3-1} k!}
{\prod\limits^{n_1+n_2+n_3-1}_{k=2n_1} k!\ \prod\limits^{n_2+n_3-n_1}_{k=1} k!}.
\label{expCone}
\ee
Next, the coefficients of the polynomials in $\hat W(\tau_i)$ appearing in \eqref{3ptfullfinal}
become
\begin{subequations}\label{leadcoeff}
\begin{align}
\hspace{-0.1cm}\frac{1}{\prod\limits_{k=n_2+1 \atop k\neq j}^{n_1+n_3+1} \sin\(\frac{\pi}{b^2}\(k-j\)\)
\prod\limits_{k=-n_2}^{n_1-n_3} \sin\(\frac{\pi}{b^2}\(k-j\)\)}
&\approx
\frac{(-1)^{n_3+j+1}\ (n_3-n_1-1+j)!\,(2 \pi \eps)^{-2n_1-1}}
{(n_2+j)!\, (j-n_2-1)!\, (n_1+n_3+1-j)!} ,
\label{simsign1}
\\
\frac{1}{\prod\limits_{k=n_2+1 }^{n_1+n_3+1} \sin\(\frac{\pi}{b^2}\(k-j\)\)
\prod\limits_{k=-n_2 \atop k\neq j}^{n_1-n_3} \sin\(\frac{\pi}{b^2}\(k-j\)\)}
&\approx
\frac{(-1)^{n_1+n_2+j+1}\ (n_2-j)!\,(2 \pi \eps)^{-2n_1-1}}
{(n_2+j)!\, (n_1-n_3-j)!\, (n_1+n_3+1-j)!}. 
\label{simsign2}
\end{align}
\end{subequations}
Thus, the $\eps$-factors from \eqref{expCone} and \eqref{leadcoeff} cancel 
and the three-point function can be written in the following form
\begin{align}
\hat C&(\beta_{n_1},\beta_{n_2},\beta_{n_3}|\tau_1,\tau_2,\tau_3)\nonumber \\
&\approx
(-1)^{\hf(n_2+n_3-n_1)(n_2+n_3-n_1-1)}\,\(\pi\over 2\)^{\sum n_i}\,
\frac{\prod\limits_{k=n_1+n_2-n_3}^{2n_2-1} k!\ \prod\limits_{k=n_1+n_3-n_2}^{2n_3-1} k!}
{\prod\limits^{n_1+n_2+n_3-1}_{k=2n_1} k!\ \prod\limits^{n_2+n_3-n_1}_{k=1} k!}
\(1+O(\eps)\)\label{3ptlim}\\
&\times
\Bigg\{\frac{1}{\hat\mu_{12}\,\hat\mu_{23}} 
\sum_{j=n_2+1}^{n_1+n_3+1} 
\frac{(-1)^{n_3+j}\ (n_3-n_1-1+j)!\,
\[{\scriptstyle \prod\limits_{k=-n_3}^{n_1-j} \left(
\hat W(\tau_1)-\hat W(\tau_3-2\pi i k)\right)
 \prod\limits_{k=1-j}^{n_2}  \left(
\hat W(\tau_2)-\hat W(\tau_3-2\pi i k)\right)}\] }
{(n_2+j)!\ (j-n_2-1)!\ (n_1+n_3+1-j)!}
\nonumber \\
&+ \frac{1}{\hat\mu_{12}\,\hat\mu_{13}}
\sum_{j=-n_2}^{n_1-n_3} 
\frac{(-1)^{n_1+n_2+j}\ (n_2-j)!\,
\[{\scriptstyle \prod\limits_{k=-n_3}^{n_1-j} \left(
\hat W(\tau_1)-\hat W(\tau_3-2\pi i k)\right)
 \prod\limits_{k=1-j}^{n_2}  \left(
\hat W(\tau_2)-\hat W(\tau_3-2\pi i k)\right)}\] }
{(n_2+j)!\ (n_1-n_3-j)!\ (n_1+n_3+1-j)!}
\Bigg\} .
\nonumber
\end{align}
According to our previous observation about the cancellation of imaginary terms,
the expression in curly brackets is a polynomial in $\hat W(\tau_i)$.
The coefficients of this polynomial are finite in the limit $b\to 1$.
Therefore, all divergences come only from the one-point functions and can be interpreted
as contact terms. To remove them, it is enough
to replace all one-point functions by their physical expressions in the $c=1$ limit,
$W_{c=1}(\tau_i)$, given in \eqref{Wphys}.
After this procedure, the resulting correlator, which we denote 
$C_{c=1}(\beta_{n_1},\beta_{n_2},\beta_{n_3}|\tau_1,\tau_2,\tau_3)$,
will be finite and possess all necessary properties of a three-point correlation function. 
In particular, one can check that it reproduces the boundary two-point function
found in the previous subsection. The details of this calculation are presented in 
appendix \ref{aptwothree}.

\section{Conclusions}
\label{concl}

In this paper we have studied the correlation functions of Liouville theory in the
limit in which the Liouville parameter $b$ goes to 1 and the central charge $c_L$ goes to 25.
The main motivation for considering this limit is that it corresponds to Liouville
theory coupled to $c=1$ matter, namely to two-dimensional string theory.

Our primary interest has been in the particular
case of two-dimensional string theory compactified on a circle at the self-dual 
radius $R=1$. This case turns out to be interesting for two reasons. First, it allows 
a very explicit representation of the correlation functions,
where all special functions characterizing the general results of Liouville theory
disappear. Second, when the Liouville momenta are restricted to the values imposed
by the compactification, additional divergences appear besides the ones that are already
present in the limit $b\to 1$. Whereas the latter divergences can be removed by a multiplicative 
renormalization of the Liouville couplings, the former are more subtle, and
some additional physical considerations were needed in order to define the theory
at these values of parameters.

Inspired by the picture emerging from the loop gas representation of non-critical strings,
we found a physical interpretation for the divergences appearing at the self-dual radius.
We associated them with various degenerate world sheet geometries, the simplest example of which is
provided by a disk with vanishing boundary length. Such geometries must be excluded from
the path integral and hence the corresponding terms, which we call contact terms, 
must be subtracted from the correlation functions. 
As a result, we proposed a precise procedure to take the $c=1$ limit based 
on the coupling of Liouville with a matter CFT representing a free scalar field 
with a background charge. 
The procedure goes as follows. After a multiplicative renormalization of the vertex operators,
one evaluates the correlation functions for special values of momenta,
takes the limit $b\to 1$, and finally subtracts all contact terms.
 
In this way we arrived at a well defined set of {\it finite} correlation functions. 
The main results are the following:

\vspace{-0.3cm}\begin{itemize}\setlength{\itemsep}{0pt}
\item \emph{three-point function on a sphere} --- \eqref{threerenR} (for a generic compactification
radius $R$);
\item \emph{bulk one-point function on a disk} --- \eqref{onecorLl} (for generic $R$)
and \eqref{physoneb} (for $R=1$);
the case of the cosmological constant operator is special and is given in 
\eqref{physonebn} (for the bulk) and in \eqref{Wphys} (for the boundary);
\item \emph{bulk-boundary two-point function} --- \eqref{nnblkbnd} (for $R=1$);
\item \emph{boundary two-point function} --- \eqref{dphys} (for $R=1$);
\item \emph{boundary three-point function} --- \eqref{3ptlim} (for $R=1$), where
in order to get the final result 
one has to perform an additional renormalization, by evaluating
the polynomial in $\hat W(\tau_i)$ and replacing the one-point functions 
by their $c=1$ limit, $W_{c=1}(\tau_i)$, given in~\eqref{Wphys}. 
\end{itemize}

\vspace{-0.3cm}

In this paper we also found several new results relevant for Liouville theory 
at generic central charge. First, we derived the boundary two-point function 
from its definition in terms of the three-point function, which is the only sensible definition
in the context of quantum gravity and string theory. As expected, 
it turned out to be proportional to the boundary reflection amplitude 
and the precise relation between them is given in~\eqref{twofunrel}.

Second, we obtained expressions for the correlation functions evaluated
at the values of Liouville momenta corresponding to degenerate momenta in the matter part.
For such momenta all special functions and integrals can be eliminated and the results
are presented in a simple form. The expressions for the bulk-boundary, two-point
and three-point boundary correlation functions can be found, respectively, in  
\eqref{Rdegen}, \eqref{Wtwobnd} and \eqref{3ptfullfinal}.  

All of these results can in principle also be found by using the techniques of matrix models.
The matrix model relevant for the $c=1$ case is matrix quantum mechanics (MQM),
which provides a solution to two-dimensional string theory both in the trivial 
and perturbed backgrounds \cite{Dijkgraaf:1992hk,Alexandrov:2002fh}
(see \cite{Alexandrov:2003ut} for a review). 
The comparison between MQM and CFT results for the one-point bulk correlation function
on a disk was already performed in various papers devoted to the study of non-perturbative effects
in two-dimensional strings. Essentially, most of the checks of the agreement between MQM
and CFT explore one or another manifestation of this correlator.

However, the correlation functions with
non-trivial boundary operators are hard to consider due to a difficulty 
in introducing such operators into the MQM framework.
Instead, they were investigated using statistical models on a fluctuating lattice 
and their loop gas representation \cite{Kazakov:1991pt,Kostov:2002uq,Kostov:2003uh,Kostov:2003cy}.
The main result of this approach
is the coincidence of some discrete equations for the correlation functions 
(similar to the shift equations used in this work) with the corresponding equations
satisfied by the CFT answers \cite{Kostov:2003uh}. 
However, also these statistical models are typically formulated for
generic values of the central charge, while the problem of the singularity of the $c=1$ limit
remains. It would therefore be quite interesting to find a model defined directly for
$c=1$ that is able to incorporate non-trivial boundary operators.

Finally, let us remark that in the case of Liouville theory on a disk
we only provided expressions for the correlation functions
with FZZT boundary conditions. The corresponding correlators for ZZ
branes~\cite{Zamolodchikov:2001ah} can be obtained by using a
known relation between FZZT and ZZ
boundary states. However, whereas a ZZ correlator in the bulk is expressed
just as a difference of FZZT correlators evaluated at special 
values of boundary parameters~\cite{Martinec:2003ka}, the case when 
non-trivial boundary operators are involved is more
complicated~\cite{Hosomichi:2001xc,Ponsot:2003ss}.
It would be nice to find a geometric interpretation for the corresponding relation,
similar to the one that holds for the bulk correlators \cite{Seiberg:2003nm,Alexandrov:2004ks}.
Our correlation functions might be useful for this purpose since they are much more
explicit than the expressions at generic central charge. 

\section*{Acknowledgments}

The authors are grateful to Valentina Petkova, Sylvain Ribault, Alexey Zamolodchikov and 
especially to Ivan Kostov and B\'en\'edicte Ponsot for very valuable discussions.


\appendix

\section{Special functions}
\label{A}

The special functions appearing in the main text are defined as follows
\begin{align}
\Gamma_b(x)&=\exp\left\{ \int\limits_0^{\infty} {dt\over t}\left[
{e^{-xt}-e^{-Qt/2}\over \(1-e^{-bt}\)\(1-e^{-t/b}\)}-{\(x-Q/2\)^2\over 2}\, e^{-t}
+{x-Q/2\over t}\right]\right\},
\label{Gamb}
\\
\Sb_b(x)&={\Gamma_b(x)\over \Gamma_b(Q-x)}
=\exp\left\{ \int\limits_0^{\infty} {dt\over t}\left[
{\sinh\((Q-2x)t\)\over 2\sinh(bt)\sinh(t/b)}+{x-Q/2\over t}\right]\right\},
\label{Sbb}
\\
\label{Ups}
\Ub_b(x)&={1\over \Gamma_b(x)\Gamma_b(Q-x)}
=\exp\left\{ \int\limits_0^{\infty} {dt\over t}\left[
\({Q\over 2}-x\)^2 e^{-t}-{\sinh^2 \({Q\over 2}-x\){t\over 2} \over
\sinh{bt \over 2}\sinh {t\over 2b} }
\right]\right\},
\end{align}
and we also use
\be
\label{Upso}
\Ub_{b,0}\mathop{=}\limits^{\rm def}\left. {d\Ub_b(x)\over dx}\right|_{x=0}=\Ub_b(b).
\ee
They possess the following properties
\beq
\Gamma_b(Q/2)&=&1, \nonumber
\\
\Gamma_b(x) &=& \Gamma_{1/b}(x),  \nonumber
\\
\Gamma_b(x+b)&=&\sqrt{2\pi}\,{b^{bx-1/2}\over \Gamma(bx)}\Gamma_b(x),  
\label{recGb}
\\
\Gamma_b(x+1/b)&=&\sqrt{2\pi}\,{b^{1/2-x/b}\over \Gamma(x/b)}\Gamma_b(x),
\nonumber 
\eeq
\bea{rclcrcl}{SU}
\Sb_b(Q/2)&=&1,
& \qquad &
\Ub_b(Q/2)&=&1,
\\
\Sb_b(x)&=&\Sb_b^{-1}(Q-x),
& \qquad &
\Ub_b(x)&=&\Ub_b(Q-x),
\\
\Sb_b(x+b)&=& 2\sin(\pi bx)\Sb_b(x),
& \qquad &
\Ub_b(x+b)&=&\gamma(bx) b^{1-2bx}\Ub_b(x),
\\
\Sb_b(x+1/b)&=& 2\sin(\pi x/b)\Sb_b(x),
& \qquad &
\Ub_b(x+1/b)&=&\gamma(x/b) b^{2x/b-1}\Ub_b(x),
\eea

\bea{rl}{analG}
\Gamma_b(x) \mbox{ is meromorphic with poles:\ }& x=-nb-m/b,\ n,m\in \Zb^{\ge 0}; 
\vspace{0.1cm}
\\
\Sb_b(x) \mbox{ is meromorphic with poles:\ }& x=-nb-m/b,\ n,m\in \Zb^{\ge 0}, 
\\
\mbox{and zeros:\ } & x=Q+nb+m/b,\ n,m\in \Zb^{\ge 0};  
\\
\Ub_b(x) \mbox{ is entire analytic with zeros:\ }& x=-nb-m/b,\ n,m\in \Zb^{\ge 0}, 
\\ 
& x=Q+nb+m/b,\ n,m\in \Zb^{\ge 0}.  \\
\eea

\section{Derivation of $U$ from $R$}
\label{B}

In this appendix we show how the one-point bulk correlation function on a disk
can be derived from the bulk-boundary correlator.
To perform this computation,
one should first take the boundary momentum to represent the boundary cosmological
constant operator, and then integrate the resulting expression with respect to $-\mub$.
After the first step the expression \eqref{bulkbnd} becomes
\begin{subequations}
\begin{align}
&\qquad R(\alpha,b|s)=  \left[\pi\mul \gamma(b^2)b^{2-2b^2}\right]^{{Q-2\alpha\over 2b}-\hf}
\CR_1\,\CR_2,
\label{bulkbndab}\\
\CR_1 &= {\Gamma_b^3\({1\over b}\)\Gamma_b\(2\alpha-b\)\Gamma_b\({2\over b}+b-2\alpha\) \over
\Gamma_b\({1\over b}+b\)\Gamma_b\({1\over b}-b\)
\Gamma_b\(b\)\Gamma_b\({1\over b}+b-2\alpha\)\Gamma_b\(2\alpha\)},
\label{R1ab}\\
\CR_2&= -2\pi i\int\limits_{-i\infty}^{i\infty}dt\, e^{-2\pi t s}
{\Sb_b\(t+\alpha-{1\over 2b}\)\Sb_b\(t-\alpha+{1\over 2b}+b\) \over
\Sb_b\(t-\alpha+{3\over 2b}+b\)\Sb_b\(t+\alpha+{1\over 2b}\)} .
\label{R2ab}
\end{align}
\end{subequations}
The first factor can be simplified to 
\be
\CR_1={b^{2\alpha\({1\over b}-b\)-1-{1\over b^2}}\over \Gamma\(1-b^2\)}\,
{\Gamma\(2\alpha b -b^2\)\over \Gamma\({1\over b^2}+1-{2\alpha\over b}\)},
\label{R1simpl}
\ee
whereas the second factor reads
\be
\CR_2= {\pi i}\int\limits_{-i\infty}^{i\infty}dt \,{e^{-2\pi t s} \over
2\sin\({\pi \over b}\(t+\alpha-{1\over 2b}\)\)  
\sin\({\pi \over b}\(t-\alpha+{1\over 2b}\)\) }  .
\label{R2simpl}
\ee
According to the integration rules described in section \ref{subsec-cordisk}, 
the integral is computed by picking residues at the following poles
\begin{equation}
	t = -\alpha+\frac{1}{2b} + \ll b\,,\qquad \ll > 0
	\qquad\qquad \text{and} \qquad\qquad
	t = \alpha- \frac{1}{2b} + \ll b\,,\qquad \ll \ge 0 .
\end{equation}
The result is
\begin{align}
\CR_2&= -{\pi b \over \sin\({\pi\over b}\(2\alpha-{1\over b}\)\)}
\(\sum_{l=1}^{\infty}e^{-2\pi s\(-\alpha+\frac{1}{2b} + \ll b\)}-
\sum_{l=0}^{\infty}e^{-2\pi s\(\alpha-\frac{1}{2b} + \ll b\)} \) 
\nonumber \\
&=
-{\pi b \over \sin\({\pi\over b}\(2\alpha-{1\over b}\)\)}\,
{\sinh\(\pi s \(2\alpha -Q\)\) \over \sinh\(\pi s b\)}.
\label{R2res}
\end{align}
Taking \eqref{bulkbndab}, \eqref{R1simpl} and \eqref{R2res} together, one obtains
\be
R(\alpha,b|s)=  -
\left[\pi\mul \gamma(b^2)\right]^{{Q-2\alpha\over 2b}-\hf}
{\Gamma\(2\alpha b -b^2\)\Gamma\({2\alpha\over b}-{1\over b^2}\) \over
b\Gamma\(1-b^2\)}\,
{\sinh\(\pi s \(2\alpha -Q\)\) \over \sinh\(\pi s b\)}.
\label{blbndab}
\ee
Finally, the integral with respect to the boundary cosmological constant gives
\begin{align}
-\int d\mub \, R(\alpha,b|s) &= 
-\pi b \sqrt{\mul\over \sin(\pi b^2)}\int ds \, \sinh(\pi s b) \, R(\alpha,b|s)
\label{finblkbnd}\\
& ={1\over \pi b}\,[\pi\mul \gamma(b^2)]^{(Q-2\alpha)/2b}
\Gamma(2b\alpha-b^2)\Gamma(2\alpha/b-1/b^2-1)\cosh\((2\alpha-Q)\pi s\).
\nonumber
\end{align}
This reproduces precisely the correlation function \eqref{onecorL}, which differs by 
a factor $2\pi$ from the standard expression used in the literature.
Its appearance can be ascribed to the residual rotation symmetry of the one-point function.

\section{Derivation of $d$ from $C$}
\label{C}

In this appendix we derive the boundary two-point function starting from 
the boundary three-point function. This is the only correct way to obtain 
it in the context of quantum gravity. The approach is the same as in all 
previous cases: to take one of the boundary operators to be the boundary 
cosmological constant operator and then integrate with respect to $-\mub$. 
However, since the expression \eqref{threepbnd} is not explicitly symmetric 
in three momenta, there are three possible ways to perform 
the calculation, all of which should give the same result. 
We are going to analyze all of the three possibilities, and this will give a non-trivial check of the
symmetry of the boundary three-point function under cyclic permutations. 

\subsection{The case $B^{s_2s_1}_{\beta_1}=B^{ss}_b$}

The first possibility is to take $\beta_1=b$, $\beta_2=\beta_3=\beta$, 
and $s_1=s_2=s$, $s_3=s'$. Then the three-point function can be represented as
\begin{subequations}
\begin{align}
&\qquad \quad C(b,\beta,\beta|s,s,s')= 
\left[\pi\mul \gamma(b^2)b^{2-2b^2}\right]^{ {1\over  2b^2}-{\beta\over b}} 
\CC_1^{(1)}\,\CC_2^{(1)}\,\CC_3^{(1)},
\label{threeone}\\
\CC_1^{(1)}& = 
{\Gamma_b\(2\beta-b\)\Gamma_b\({2\over b}+b-2\beta\)
\Gamma_b^2\({1\over b}\)
\over \Gamma_b\({1\over b}+b\)\Gamma_b\({1\over b}-b\)\Gamma_b^2\({1\over b}+b-2\beta\)}
={ b^{2\beta\({1\over b}-b\)-{1\over b^2}}  \over \Gamma\(1-b^2\)}\,
{\Gamma\(2\beta b-b^2 \)\over \Gamma\({1\over b^2}+1-{2\beta \over b}\)}\,\Sb_b(2\beta),
\label{CC1one} \\
\CC_2^{(1)}&=  
{\Sb_b\({1\over b}+b-\beta+i{s-s'\over 2}\)\Sb_b\({1\over b}+b-\beta-i{s+s'\over 2}\)
\over \Sb_b\(\beta+i{s-s'\over 2}\) \Sb_b\(\beta-i{s+s'\over 2}\)},
\label{CC2one}\\
\CC_3^{(1)}&= -4\pi i
\int\limits_{-i\infty}^{i\infty}dt\, \textstyle
{\Sb_b\({1\over b}+i{s}+t\)\Sb_b\({1\over b}+t\)
\Sb_b\(\beta+i{s-s'\over 2}+t\) \Sb_b\({1\over b}+b-\beta+i{s-s'\over 2}+t\)
\over \Sb_b\({2\over b}+b-\beta+i{s-s'\over 2}+t\) 
\Sb_b\({1\over b}+\beta+i{s-s'\over 2}+t\)
\Sb_b\({1\over b}+b+i{s}+t\) \Sb_b\({1\over b}+b+t\)}
\nonumber \\
&={\pi i\over 4}\int\limits_{-i\infty}^{i\infty}
{dt\over \sin(\pi b t) \sin[\pi b (is+t)] \sin\[{\pi \over b}\(\beta+i{s-s'\over 2}+t\)\]
\sin\[{\pi \over b}\({1\over b}-\beta+i{s-s'\over 2}+t\)\]}.
\label{CC3one}
\end{align}
\end{subequations}
Note that, in contrast to the integral appearing in the expression for the bulk-boundary correlator,
the integrand in \eqref{CC3one} does not vanish when $t\to \pm \infty$.
Therefore, in principle, we cannot close the integration contour 
either in the left or in the right half plane. Nevertheless, let us assume that 
it is possible and the contour is closed in the right half plane.
We will justify this assumption shortly.
Then, using the integration rules described in section \ref{subsec-cordisk}, 
the last factor can be evaluated by residues. This gives the following result
\begin{equation}\label{CC3onesm}
\begin{gathered}
\CC_3^{(1)} = {\pi \over 2}\sum\limits_{k=0}^{\infty}
\textstyle\[ 
{1\over \sin(\pi i s b)}\(
{1 \over 
\sin\[{\pi \over b}\({k\over b}+\beta+i{s-s'\over 2}\)\]
\sin\[{\pi \over b}\({k+1\over b}-\beta+i{s-s'\over 2}\)\]}
-{1\over 
\sin\[{\pi \over b}\({k\over b}+\beta-i{s+s'\over 2}\)\]
\sin\[{\pi \over b}\({k+1\over b}-\beta-i{s+s'\over 2}\)\]} \)
\right. \\
 +  \left. 
\textstyle {b\over \sin\[{\pi\over b}\(2\beta-{1\over b}\)\]}\( 
{1 \over \sin\[\pi b\(kb +\beta-i{s-s'\over 2}\)\] 
\sin\[\pi b\( kb +\beta+i{s+s'\over 2}\)\]}
-{1\over  
\sin\[\pi b\( (k+1)b-\beta-i{s-s'\over 2}\)\]
\sin\[\pi b\( (k+1)b -\beta+i{s+s'\over 2}\)\]} \)\].
\end{gathered}
\end{equation}
It is easy to convince oneself that this expression can be rewritten in the following form
\begin{equation}\label{CC3shift} 
\begin{split}
\CC_3^{(1)} & =
\textstyle{\pi\over 2i \sinh(\pi s b)\,\sin\[{\pi\over b}\(2\beta-{1\over b}\)\]}\\
&\quad\times\scriptstyle\Big[ {1/b\over 1-e^{-{2i\over b} \p_s}}\(
\cot\[{\pi \over b}\({1\over b}-\beta+i{s-s'\over 2}\)\]
-\cot\[{\pi \over b}\(\beta+i{s+s'\over 2}\)\]
-\cot\[{\pi \over b}\(\beta+i{s-s'\over 2}\)\]
+\cot\[{\pi \over b}\({1\over b}-\beta+i{s+s'\over 2}\)\] \)  \\
&\quad+\,
\scriptstyle {b\over 1-e^{-2i b \p_s}}\( 
\cot\[\pi b\(b -\beta+i{s-s'\over 2}\)\] 
-\cot\[\pi b\(\beta+i{s+s'\over 2}\)\]
-\cot\[\pi b\( \beta+i{s-s'\over 2}\)\]
+\cot\[\pi b\( b -\beta+i{s+s'\over 2}\)\] \)\Big], 
\end{split}
\end{equation}
where we used a shift operator in $s$ similar to the one in \eqref{shift}. 
 
Let us consider the function defined by 
\be
\Tb_b(x)\equiv
{b \over 1-e^{b\p_x}} \cot(\pi b x) +{1/b\over 1-e^{{1\over b}\p_x}} \cot(\pi x / b). 
\label{defder}
\ee
Applying the operators $1-e^{b\p_x}$ and $1-e^{{1\over b}\p_x}$ to this definition,
one can establish the following properties under shifts of the argument
\beq
\Tb_b(x+b)&=&\Tb_b(x)-b\,\cot(\pi b x),
\\
\Tb_b(x+1/b)&=&\Tb_b(x)-{1\over b}\,\cot(\pi x/b).
\eeq
On the other hand, as follows from \eqref{SU}, 
the function $-{1\over \pi}\,{d\over dx}\log \Sb_b(x)$ satisfies exactly the same properties.
This allows to conclude (for non-rational $b$) that
\be
\Tb_b(x)=-{1\over \pi}\,{d\over dx}\log \Sb_b(x)+c_0,
\label{relTS}
\ee
where the constant $c_0$ can in principle be found, but 
in any case our results are independent of it.

From the property \eqref{relTS}, it is clear that \eqref{CC3shift} can be 
represented as
\be
\CC_3^{(1)} =
{1\over \sinh(\pi s b)\,\sin\[{\pi\over b}\(2\beta-{1\over b}\)\]}\,
{\p\over \p s} \log 
{\Sb_b\({1\over b}+b-\beta+i{s-s'\over 2}\)\Sb_b\({1\over b}+b-\beta+i{s+s'\over 2}\)
\over \Sb_b\(\beta+i{s+s'\over 2}\) \Sb_b\(\beta+i{s-s'\over 2}\)},
\label{CC3S}
\ee
so that the undetermined constant $c_0$ does not enter the result.
The argument of the logarithm coincides in fact with the factor $\CC_2^{(1)}$ in~\eqref{CC2one}.
Therefore, the full three-point function can be rewritten as
\be
C(b,\beta,\beta|s,s,s')= 
 \left[\pi\mul \gamma(b^2)b^{2-2b^2}\right]^{ {1\over  2b^2}-{\beta\over b}} 
{ b^{2\beta\({1\over b}-b\)-{1\over b^2}}  \over \Gamma\(1-b^2\)}\,
{\Gamma\(2\beta b-b^2 \)\Gamma\({2\beta \over b}-{1\over b^2}\)\over 
\pi \sinh(\pi s b)}\,\Sb_b(2\beta)\ \frac{\partial \CC_2^{(1)}}{\partial s}.
\label{fullCC}
\ee
Using the fact that 
\be
\Sb_b(2\beta)=
{2\pi \, b^{2\beta\(b-{1\over b}\)-b^2+{1\over b^2}+1}\over 
\Gamma\(2\beta b-b^2\)\Gamma\({2\beta \over b}-{1\over b^2}-1\)}\,
{\Gamma_b(2\beta-Q)\over \Gamma_b(Q-2\beta)} 
\label{SbGammab}
\ee
and rewriting the derivative in terms of the boundary cosmological constant, one finds
\be
C(b,\beta,\beta|s,s,s')= 
{\p\over \p \mub}\[
{2(2\beta-Q)
\left[\pi\mul \gamma(b^2)b^{2-2b^2}\right]^{(Q-2\beta)/2b}
\Gamma_b(2\beta-Q)\Gamma_b^{-1}(Q-2\beta)
\over \Sb_b\(\beta+i{s+s'\over 2}\)\Sb_b\(\beta-i{s+s'\over 2}\)
\Sb_b\(\beta+i{s-s'\over 2}\)\Sb_b\(\beta-i{s-s'\over 2}\) }\] .
\label{fullCC2}
\ee
Thus, the integral with respect to $-\mub$ can be trivially evaluated and one
obtains a result proportional to the boundary reflection amplitude \eqref{twopbnd},
as shown in \eqref{twofunrel}.
Note that the proportionality factor is very similar to the one 
appearing in the relation between the corresponding bulk quantities 
\eqref{difreftwo}.

Let us now return to the discussion of the integral $\CC^{(1)}_3$ in~\eqref{CC3one}. 
Although, as we discussed above, our procedure to compute it was not rigorous,
the result that we get for the two-point function is perfectly sensible. 
Therefore, one may wonder if there is an 
alternative way to justify the computation.
For this we will consider $\CC^{(1)}_3$ as an analytic function of its parameters,
the boundary variables $s,\ s'$ and momentum $\beta$, and 
derive some difference equations satisfied by the integral.
Such equations can be obtained by shifting the parameters in various ways. 
In principle, at least for generic $b$, these ``shift equations'' should fix 
the solution completely. Then we prove that the equations we derived 
are satisfied by the answer coming from our computation by residues.

This strategy will turn out to be extremely useful in other 
cases as well: in the next part of this appendix, we will use it in order 
to perform the computation of the two-point function, starting from the three-point 
function, in the case in which $\beta_2=b$ or $\beta_3=b$. In appendix~\ref{secshifts}, 
the same strategy will allow us to obtain the integration rules for the 
three-point function that we use in the main text.

\begin{figure}
\begin{center}
\includegraphics[scale=0.65]{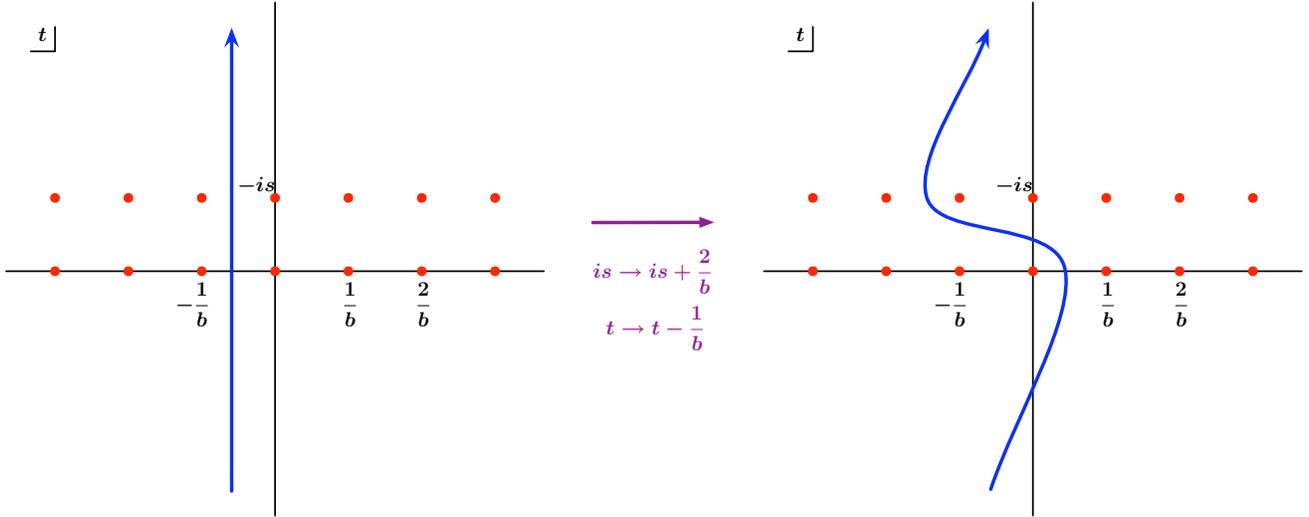}
\caption{{\small Shift of the integration contour.}}
\label{f:shift}
\end{center}
\end{figure}
To find some shift equations for the integral \eqref{CC3one}, it is useful to rewrite it as
\be
\sinh(\pi b s)\,\CC_3^{(1)}={\pi\over 4}\int\limits_{-i\infty}^{i\infty}dt\,
{\cot(\pi b t) -\cot[\pi b (is+t)]\over  
\sin\[{\pi \over b}\(\beta+i{s-s'\over 2}+t\)\]
\sin\[{\pi \over b}\({1\over b}-\beta+i{s-s'\over 2}+t\)\]}.
\label{CC3onere}
\ee
Now let us consider, for instance, the shift $is\to is +2/b$. It has an effect only on 
the arguments of the sine functions in the denominator. However, the change of the arguments 
can be compensated by shifting the integration variable $t\to t-1/b$. One could think that
this brings us back to the initial integral, but this is not precisely the case. The difference
comes from the definition of the contour of integration described in section
\ref{subsec-cordisk}. Despite our shifts not being seen in the cotangent functions
in the numerator, it is easy to see that they modify the integration contour in such a way 
that one pole of $\cot(\pi b t)$ passes the contour from the right to the left and one pole
of $\cot[\pi b (is+t)]$ from the left to the right.
We show this schematically in fig. \ref{f:shift}.
Therefore, in order to revert to the original integral, one has to add (subtract) the contribution
of these poles. As a result, one arrives at the following equation satisfied by
the integral \eqref{CC3onere}  
\begin{equation}
\begin{split}
&\(e^{-{2i\over b}\p_s}-1\)\[\sinh(\pi b s)\,\CC_3^{(1)}\]\\
&\qquad=\textstyle
{\pi i\over 2b} \({1\over  
\sin\[{\pi \over b}\(\beta+i{s-s'\over 2}\)\]
\sin\[{\pi \over b}\({1\over b}-\beta+i{s-s'\over 2}\)\]} +
{1\over  \sin\[{\pi \over b}\({1\over b}-\beta+i{s+s'\over 2}\)\]
\sin\[{\pi \over b}\(\beta+i{s+s'\over 2}\)\]} \)\\
&\qquad=\, \scriptstyle
{\pi i\over 2b\sin\[{\pi\over b}\(2\beta-{1\over b}\)\]  }
\(\cot\[{\pi \over b}\({1\over b}-\beta+i{s-s'\over 2}\)\]
-\cot\[{\pi \over b}\(\beta+i{s-s'\over 2}\)\]
-\cot\[{\pi \over b}\(\beta+i{s+s'\over 2}\)\]
+\cot\[{\pi \over b}\({1\over b}-\beta+i{s+s'\over 2}\)\] \).
\label{deqCC3one}
\end{split}
\end{equation}

The shift $is\to is+2b$ works in a slightly different way. 
It does not influence the denominator and changes only the argument of the second cotangent.
Because of this, it is convenient to consider \eqref{CC3onere} as a sum of two integrals
and shift the integration variable, $t\to t-2b$, only in the second one.
Then again the resulting integrals differ from the initial ones only
by the contours of integration which pick up additional poles coming from the sine functions
in the denominator. Considerations similar to the previous ones then lead to the following equation
\begin{multline}
\(e^{-{2ib}\p_s}-1\)\[\sinh(\pi b s)\,\CC_3^{(1)}\]\\
=\, \scriptstyle
{\pi ib\over 2\sin\[{\pi\over b}\(2\beta-{1\over b}\)\]  }
\(\cot\[\pi b\(b -\beta+i{s-s'\over 2}\)\] 
-\cot\[\pi b\( \beta+i{s-s'\over 2}\)\]
-\cot\[\pi b\(\beta+i{s+s'\over 2}\)\]
+\cot\[\pi b\( b -\beta+i{s+s'\over 2}\)\] \).
\label{deqbCC3one}
\end{multline}

In the same way, one can find other difference equations which involve shifts
of $s'$ and $\beta$. It is easy to check that \eqref{deqCC3one}, \eqref{deqbCC3one},
as well as these other equations are satisfied by the function given in \eqref{CC3S}.
Thus, the study of shift equations yields
the same result that we reach by summing residues.  
This completes the analysis of the first of the three ways to derive the two-point 
function.

\subsection{The cases $B^{s_3s_2}_{\beta_2}=B^{ss}_b$ and 
$B^{s_1s_3}_{\beta_3}=B^{ss}_b$}
\label{aptwoother}

Now we turn to the computation of the two-point function from the three-point 
correlator in the two remaining cases. The first one is obtained by taking
$\beta_2=b$, $\beta_1=\beta_3=\beta$ and $s_2=s_3=s$, $s_1=s'$, whereas the second case
corresponds to the choice $\beta_3=b$, $\beta_1=\beta_2=\beta$ and $s_1=s_3=s$, $s_2=s'$. 
However, one can show that the expressions for the three-point function arising
after these substitutions are related by a simple change of variables
\be
s\to -s, \qquad s'\to -s', \qquad t\to \beta+i{s-s'\over 2}-Q-t,
\label{changesss}
\ee
where we also showed the change of the integration variable that is needed.
Since the final result, which is expected to coincide with \eqref{fullCC2}, is 
an even function of the boundary parameters, this fact ensures that 
the two possibilities we consider here will give the same two-point function.
Therefore, we concentrate only on the second case, $\beta_3=b$. 

With this choice of the parameters, the three-point function reduces to:
\begin{subequations}
\begin{equation}
C(\beta,\beta,b|s,s',s)= 
\left[\pi\mul \gamma(b^2)b^{2-2b^2}\right]^{ {1\over  2b^2}-{\beta\over b}} 
\CC^{(3)}_1\,\CC^{(3)}_2\,\CC^{(3)}_3,
\label{threethree}
\end{equation}
\begin{align}
\CC^{(3)}_1& = 
{\Gamma_b\(b\)\Gamma_b\({2\over b}+b-2\beta\)
\Gamma_b\({1\over b}+2b-2\beta\)\Gamma_b\({1\over b}\)
\over \Gamma_b\({1\over b}+b\)\Gamma_b^2\({1\over b}+b-2\beta\)
\Gamma_b\({1\over b}-b\)} 
=\frac{b\, \CC^{(1)}_1}{\Sb_b\(2\beta-b\)} ,
\label{CC1three} \\
\CC^{(3)}_2&=  
{\Sb_b\({1\over b}\)\Sb_b\({1\over b}-i{s}\)
\over \Sb_b\(\beta+i{s'-s\over 2}\) \Sb_b\(\beta-i{s'+s\over 2}\)} ,
\label{CC2three} \\
\CC^{(3)}_3&= -4 \pi i
\int\limits_{-i\infty}^{i\infty}dt\, \textstyle
{\Sb_b\({1\over b}+b-\beta+i{s+s'\over 2}+t\)
\Sb_b\({1\over b}+b-\beta-i{s-s'\over 2}+t\) 
\Sb_b\(\beta+i{s'-s\over 2}+t\) 
\Sb_b\({1\over b}+b-\beta+i{s'-s\over 2}+t\)
\over
\Sb_b\({1\over b}+b+i{s'}+t\) \Sb_b\({1\over b}+b+t\)
\Sb_b\({2\over b}+b-\beta+i{s'-s\over 2}+t\) 
\Sb_b\({1\over b}+2b-\beta+i{s'-s\over 2}+t\)}
\nonumber \\
&= -\pi i \int\limits_{-i\infty}^{i\infty} \textstyle
{dt \over \sin\[\pi b\(b-\beta-i{s-s'\over 2}+t\)\]
\sin\[{\pi\over b}\({1\over b}-\beta-i{s-s'\over 2}+t\)\]}
{\Sb_b\(\beta+i{s'-s\over 2}+t\) \Sb_b\({1\over b}+b-\beta+i{s+s'\over 2}+t\)
\over \Sb_b\({1\over b}+b+t\)\Sb_b\({1\over b}+b+i{s'}+t\)} .
\label{CC3three}
\end{align}
\end{subequations}

By shifting the parameters in $\CC^{(3)}_3$ in various ways, one can again find 
some shift equations satisfied by the integral \eqref{CC3three}. Due to the
symmetry of the integral under $b\to 1/b$, it is enough to find 
equations which use only shifts by multiples of $b$. Then the symmetry guarantees 
the existence of similar equations with $b$ replaced by $1/b$, and the final result must 
be symmetric as well.
We find for instance the following three equations 
\begin{subequations}
\begin{align}
	\left[\frac{\sin\left[ \pi b \left(\beta - i \frac{s-s'}{2} 
	\right) \right] e^{-2 i b \p_{s'}}
	- \sin\left[ \pi b \left(b + i s' \right) \right] e^{-i b (\p_{s}+\p_{s'})}}
	{\sin\left[ \pi b \left(\beta - b - i \frac{s+s'}{2} \right) \right]}
	-1\right]\ \CC_3^{(3)} &= 0 ,\\
	\left[\frac{\sin\left[ \pi b \left(\beta -b - i \frac{s+s'}{2} \right) \right] 
	e^{-2 i b \p_{s'}}
	- \sin\left[ \pi b \left(b + i s' \right) \right] 
	e^{\frac{b}{2}\p_{\beta} - i b \p_{s'}}}
	{\sin\left[ \pi b \left(\beta - i \frac{s-s'}{2} \right) \right]}
	-1\right]\ \CC_3^{(3)} &= 0 ,
\end{align}
\begin{multline}\label{shift3three}
	\textstyle
	\Bigg[\frac{\sin\left[ \pi b \left( \beta- b - i \frac{s+s'}{2} \right) \right]}
	{\sin\left[ \pi b \left(b - 2 \beta + i s \right) \right]}	
	\( e^{-\frac{b}{2} \p_{\beta}- i b \p_{s'}}
	+\frac{\sin\left[ \pi b \left( i s - b \right) \right]}
	{\sin\left[ \pi b \left( 2 \beta - b \right) \right]}\, e^{ib \(\p_{s}-\p_{s'}\)} \) 
	- \frac{\sin\left[ \pi b \left( \beta + i \frac{s+s'}{2} \right) \right]}
	{\sin\left[ \pi b \left( 2 \beta - b \right) \right]}\, 
	e^{\frac{b}{2}\p_{\beta} - i b \p_{s'}}
	-1\Bigg]\CC_3^{(3)} \\
	\textstyle
	=\frac{2\pi b\, \sin\left[ \pi b \left(b - \beta + i \frac{s+s'}{2} \right) \right]}
	{\sin\left[ \pi b \left( \beta + i \frac{s+s'}{2} \right) \right]
	\sin\left[ \pi b \left( 2 \beta - b \right) \right]}\,
	\frac{\Sb_b\(2\beta-b-\frac{1}{b}\) \Sb_b\(i s\)}
	{\Sb_b\(\beta+i{s+s'\over 2}\) \Sb_b\(\beta+i{s-s'\over 2}\)} . 
\end{multline}
\end{subequations}
The origin of the non-homogeneous right hand side in the third equation 
can be again traced back to the shift of the integration contour. Both shift operators 
$e^{-\frac{b}{2} \partial_{\beta}- i b \partial_{s'}}$ and 
$e^{ib \(\partial_{s}-\partial_{s'}\)}$ transform the arguments of the
sine functions in the integrand~\eqref{CC3three} 
so that one has to pick the residues at the poles situated at 
$t = i\frac{s-s'}{2}+\beta-2b$ and $t = i\frac{s-s'}{2}+\beta-b-\frac{1}{b}$ 
in order to revert to the original contour of integration. 
The sum of the contributions of these four residues gives the right hand side of~\eqref{shift3three}.

One can show that the following expression for $\CC_3^{(3)}$
\begin{equation}\label{CC3S3}
	\CC_3^{(3)} = -
	\frac{4i\,\Sb_b\(2\beta-b-\frac{1}{b}\) \Sb_b\(i s\)}
	{\Sb_b\(\beta+i{s+s'\over 2}\) \Sb_b\(\beta+i{s-s'\over 2}\)}\
	\frac{\partial}{\partial s} \log 
	{\Sb_b\({1\over b}+b-\beta+i{s-s'\over 2}\)\Sb_b\({1\over b}+b-\beta+i{s+s'\over 2}\)
	\over \Sb_b\(\beta+i{s+s'\over 2}\) \Sb_b\(\beta+i{s-s'\over 2}\)} .
\end{equation}
satisfies all of the above shift equations. When combined with~\eqref{CC1three} 
and~\eqref{CC2three}, \eqref{CC3S3} reproduces precisely the expression~\eqref{fullCC} 
and thus leads to the same two-point function that was obtained in the first part 
of this appendix.

We conclude that all three possibilities to get the two-point function starting from 
the three-point correlator give the same result \eqref{twofunrel}, which differs from the 
boundary reflection amplitude by a simple momentum dependent factor. This 
provides a non-trivial check of the cyclic symmetry of the representation \eqref{threepbnd} 
for the three-point function and gives strong evidences in favor of its correctness.
 
\subsection{One-point functions}
\label{ap-onepoint}

We can also derive the one-point function of the cosmological constant
operator starting from the result \eqref{twofunrel}. Putting $\beta=b$ and $s_1=s_2=s$,
one obtains 
\beq
d(b|s,s) & =& 2\({1\over b}-b\) 
\left[\pi\mul \gamma(b^2)b^{2-2b^2}\right]^{{1-b^2\over 2b^2}}
{\Gamma_b\(b-{1\over b}\)\over \Gamma_b\({1\over b}-b\)}\, 
{\Sb_b\(is+{1\over b}\)\over \Sb_b\(is+b\)}\,
{\Sb_b\({1\over b}\)\over \Sb_b\(b\)}
\nonumber  \\
& =& -{2\over b }
\left[\pi\mul \gamma(b^2)\right]^{{1-b^2\over 2b^2}}
{\Gamma\(2-{1\over b^2}\)\over \Gamma\(1-b^2\)}\,
{\sinh(\pi s/b)\over \sinh(\pi s b)}.
\label{twobndsim}
\eeq
The first integral with respect to $-\mub$ gives the boundary one-point function of $B_b^{ss}$
given in \eqref{oneW}, which we denote by $W(s)$ since it corresponds to the resolvent in the
matrix model formulation.
The second integral gives the disk partition function with FZZT boundary conditions
\be
Z(s)=-\left[\pi\mul \gamma(b^2)\right]^{{1+b^2\over 2b^2}}
{b^2\Gamma\(2-{1\over b^2}\) \Gamma\(1-b^2\)\over \pi^2} 
\({\cosh\(\(b+{1\over b}\)\pi s\)\over b+{1\over b}}+
{\cosh\(\(b-{1\over b}\)\pi s\)\over b-{1\over b}}\).
\ee 
As a last step one has to evaluate the derivative with respect to $-\mu$, keeping
$\mub$ constant. The result
\be
-{\p Z\over \p\mu}={1\over \pi b}
\left[\pi\mul \gamma(b^2)\right]^{{1-b^2\over 2b^2}}
\Gamma\(b^2\)\Gamma\(1-\textstyle{1\over b^2}\) 
\cosh\( \(b-\textstyle{1\over b}\)\pi s\) = U(b|s)
\label{derZdisk}
\ee
shows that the one-point function obtained from the boundary three-point function, 
with the normalization that we have chosen, precisely
coincides with the same quantity derived from the bulk-boundary correlator.
This justifies the inclusion of the factor $4\pi$ into the initial expression
for the three-point function.

\section{Details on the boundary three-point function}

\subsection{Evaluating the integral}
\label{secshifts}

In this appendix we study the integral \eqref{C32} appearing in the calculation
of the three-point function for degenerate momenta of the matter part. 
For convenience we rewrite it here: 
\begin{equation}\label{C32repeat}
\begin{split}
\mathcal{C}_3 =&\,
\frac{\pi i(-1)^{\sum n_i}}{4^{n_1}} \int\limits_{-i\infty}^{i\infty}dt\, 
\frac{\Sb_b\(t+{n_1+1\over b}+i{\tau_1+\tau_2\over 2\pi b}\)\, 
\Sb_b\(t+{n_1+1\over b}-i{\tau_1-\tau_2\over 2\pi b}\)}
{\Sb_b\(t+{n_2+1\over b}+i{\tau_2-\tau_3\over 2\pi b}\)\, 
\Sb_b\(t+b+{1\over b}\)}
\\
&\times 
\textstyle
\[\prod\limits_{k=n_2+1}^{n_1+n_3+1} \sin\left(\frac{\pi}{b}\left(t+i\frac{\tau_2-\tau_3}{2\pi b}
+\frac{k}{b}\right)\right)
\prod\limits_{k=-n_2}^{n_1-n_3} \sin\left(\frac{\pi}{b}\left(t+i\frac{\tau_2-\tau_3}{2\pi b}
+\frac{k}{b}\right)\right)\]^{-1} .
\end{split}
\end{equation}
First, following the method used in appendix \ref{C}, we derive some shift equations 
satisfied by this integral. They can be obtained by applying various shift operators
to the integrand and, if necessary, also by shifting the integration variable.
As explained in detail in appendix \ref{C}, one should take a special care 
of a possible change of the integration contour which can produce non-homogeneous
contributions to the equations. In this way we find the following equations%
\footnote{There are additional equations one can write besides~\eqref{mightnot},
but we think that the ones presented here are already non-trivial enough 
to illustrate the correctness of the result we are going to propose.}
\begin{subequations}\label{mightnot}
\begin{align}
\[ \frac{\sinh\({\tau_1-\tau_2\over 2b^2}-{\pi i n_1\over b^2}\) \, e^{-2\pi i \p_{\tau_2}}
-\sinh\({\tau_2\over b^2}-{\pi i \over b^2}\)\, e^{-\pi i \(\p_{\tau_1}+\p_{\tau_2}+\p_{\tau_3}\)}}
{\sinh\({\tau_1+\tau_2\over 2b^2}-{\pi i (n_1+1)\over b^2}\)} -1\]\CC_3 &= 0,
\label{eqC31b}
\\
\[ \frac{\sinh\({\tau_1-\tau_2\over 2}+\pi i b^2\) \, e^{-2\pi i b^2\p_{\tau_2}}
+(-1)^{n_1}\sinh\(\tau_2-\pi i b^2\)\, e^{-\pi i b^2\(\p_{\tau_1}+\p_{\tau_2}+\p_{\tau_3}\)}}
{\sinh{\tau_1+\tau_2\over 2}} -1\]\CC_3 &= 0,
\label{eqC31}
\end{align}
\be
\[e^{-2\pi i b^2\p_{\tau_3}}-1\]\CC_3=
2\pi i \(\sum\limits_{j=-n_2}^{n_1-n_3}\res\limits_{t=t_{j0}}+
\sum\limits_{j=n_2+1}^{n_1+n_3+1}\res\limits_{t=t_{j1}}\)\mathcal{I}(t),
\label{eqC32}
\ee
\begin{align}
&\[ \frac{\sinh\({\tau_1-\tau_2\over 2}-\pi i b^2\)}{\sinh{\tau_1+\tau_2\over 2}}
\(e^{-2\pi i b^2 \p_{\tau_1}}+\frac{\sinh(\tau_1-\pi i b^2)}{\sinh(\tau_2-\pi i b^2)}\, 
e^{-2\pi i b^2 \p_{\tau_2}}\) 
-\frac{\sinh(\tau_1-\pi i b^2)}{\sinh(\tau_2-\pi i b^2)}+1\]\CC_3 &
\nonumber
\\
&\ \ = -2\pi i\, \frac{\sinh\(\tau_1-\pi i b^2\)}{\sinh{\tau_1+\tau_2\over 2}}
\(\sum\limits_{j=-n_2}^{n_1-n_3}\res\limits_{t=t_{j0}}+
\sum\limits_{j=n_2+1}^{n_1+n_3+1}\res\limits_{t=t_{j1}}\)
\frac{\sin(\pi b t)\,\mathcal{I}(t)}{\sin\[\pi b\(t-i{\tau_1-\tau_2\over 2\pi b}-b\)\]} ,&
\label{eqC33}
\end{align}
\end{subequations}
where $\mathcal{I}(t)$ denotes the integrand (together with the numerical prefactor)
and $t_{jl}$ are its poles coming from the sine functions in the denominator
\be
t_{j\ll} = - i\frac{\tau_2-\tau_3}{2\pi b} - \frac{j}{b} + \ll b .
\label{C3poles}
\ee
The corresponding residues of $\mathcal{I}(t)$ were calculated in \eqref{res1}.
In the last equation they appear multiplied by the factor 
$\frac{\sin(\pi b t_{jl})}{\sin\[\pi b\(t_{jl}-i{\tau_1-\tau_2\over 2\pi b}-b\)\]}$.

Then, by straightforward but tedious calculations, one can show that
all equations are satisfied by the following function
\begin{equation}
\begin{split}
\mathcal{C}_3  &= 2 \pi b\ (-1)^{\sum n_i+1} 4^{n_2+n_3}\,\muc\,
{\Sb_b\(b-{n_3\over b}-i{\tau_1-\tau_3\over 2\pi b}\)
\Sb_b\(b-{n_3\over b}+i{\tau_1+\tau_3\over 2\pi b}\)
\over \Sb_b\({n_2+1\over b}-i{\tau_2-\tau_3\over 2\pi b}\)
\Sb_b\({n_2+1\over b}+i{\tau_2+\tau_3\over 2\pi b}\)}
\\
&\qquad\qquad \times \Bigg\{\frac{1}{\hat\mu_{12}\hat \mu_{23}}
\sum_{j=n_2+1}^{n_1+n_3+1} 
\frac{\mathcal{A}_{j}}
{\prod\limits_{k=n_2+1 \atop k\neq j}^{n_1+n_3+1} \sin\(\frac{\pi}{b^2}\(k-j\)\)
\prod\limits_{k=-n_2}^{n_1-n_3} \sin\(\frac{\pi}{b^2}\(k-j\)\)}
\\
&\qquad\qquad\quad + \frac{1}{\hat\mu_{12}\hat \mu_{13}}
\sum_{j=-n_2}^{n_1-n_3} 
\frac{\mathcal{A}_{j}}
{\prod\limits_{k=n_2+1 }^{n_1+n_3+1} \sin\(\frac{\pi}{b^2}\(k-j\)\)
\prod\limits_{k=-n_2 \atop k\neq j}^{n_1-n_3} \sin\(\frac{\pi}{b^2}\(k-j\)\)}\Bigg\},
\end{split}
\label{exprC3}
\end{equation}
where $\CA_j$ is given in \eqref{defAA} or in \eqref{WAA}. Since the ratio of $\Sb_b$ functions
coincides with $\CC_2^{-1}$ (cf. \eqref{C2}), this is precisely the result \eqref{3ptfinal}
which we obtained in the main text by closing the contour
of integration in \eqref{C32repeat} in the right half plane and summing
the residues of the integrand only at the poles which are due to the sine functions.
This gives convincing evidence of the correctness of the use of these ``effective integration rules".

\subsection{Explicit results for the three-point function}
\label{explicthree}

This subsection contains explicit expressions of the three-point function for various 
low values of momenta $n_i$. We consider all cases up to
$\sum n_i\le 4$ and satisfying the triangle inequality $|n_2- n_3|\le n_1 \le n_2+ n_3$.
The results are expressed in terms of the one point functions 
$\hat W_i\equiv \hat W(\tau_i)$. We will use the shorthand $\muc_i = \mubc (\tau_i)$, 
and we recall that the expression of the coefficients $a_n$ is given in~\eqref{const_c}.

\begin{align}
	\hat{C}&(\beta_0,\beta_0,\beta_0|\tau_1,\tau_2,\tau_3)
		= \frac{1}{\muc_{12}\ \muc_{23}\ \muc_{31}}
		\left[ \muc_1 (\hw_2-\hw_3)+ \muc_2 (\hw_3-\hw_1)+\muc_3 (\hw_1-\hw_2) \right]\,,\\
		\ \nonumber \\
	\hat{C}&(\beta_0,\beta_1,\beta_1|\tau_1,\tau_2,\tau_3)
		= \frac{a_1}{\muc_{12}\ \muc_{23}\ \muc_{31}}\nonumber\\
		&\times \Bigg\{
		\left[ \muc_1 \left( \hw_2^3 - \hw_3^3
		- \frac{\sin{\frac{3\pi}{b^2}}}{\sin{\frac{\pi}{b^2}}} \hw_2 \hw_3 (\hw_2 - \hw_3)
		-\frac{16 b^2}{\pi^2} \muc^{\frac{1}{b^2}} \cos^2{\textstyle\frac{\pi}{b^2}} (\hw_2-\hw_3)
		\right) - (1 \leftrightarrow 2) \right]\\
		&+\muc_3 \left( \hw_1^3 - \hw_2^3
		+ \frac{\sin{\frac{3\pi}{b^2}}}{\sin{\frac{\pi}{b^2}}}
		(\hw_3^2 (\hw_1 - \hw_2) - \hw_3 (\hw_1^2-\hw_2^2))
		-\frac{16 b^2}{\pi^2} \muc^{\frac{1}{b^2}} \cos^2{\textstyle\frac{\pi}{b^2}} (\hw_1-\hw_2)\right)
		\Bigg\}\nonumber\\
		\ \nonumber \\
	\hat{C}&(\beta_1,\beta_1,\beta_1|\tau_1,\tau_2,\tau_3)
		= \frac{a_{3/2}}{\muc_{12}\ \muc_{23}\ \muc_{31}}
		\frac{2 \cos{\textstyle\frac{2\pi}{b^2}}\ b^2\ \Gamma\(\frac{2}{b^2}\)^2}
		{\Gamma\(\frac{1}{b^2}\) \Gamma\(\frac{4}{b^2}-1\)}
		\Bigg\{
		\muc_1 \Bigg( \hw_2^4 - \hw_3^4\nonumber\\
		&- 2 \cos{\textstyle\frac{2\pi}{b^2}}
		( \hw_1 (\hw_2^3 - \hw_3^3) + \hw_2 \hw_3 (\hw_2^2 - \hw_3^2))
		+ 2 \cos{\textstyle\frac{2\pi}{b^2}}\ \frac{\sin{\frac{3\pi}{b^2}}}{\sin{\frac{\pi}{b^2}}}
		\hw_1 \hw_2 \hw_3 (\hw_2 - \hw_3)\\
		&- \frac{16 b^2}{\pi^2} \muc^{\frac{1}{b^2}} \cos^2{\textstyle\frac{\pi}{b^2}}
		\( \hw_2^2-\hw_3^2 - 2 \cos{\textstyle\frac{2\pi}{b^2}} \hw_1 (\hw_2 - \hw_3)
		\)\Bigg)+\text{cyclic permutations}\Bigg\}\nonumber\\
		\ \nonumber \\
	\hat{C}&(\beta_0,\beta_2,\beta_2|\tau_1,\tau_2,\tau_3)
		= \frac{a_2}{\muc_{12}\ \muc_{23}\ \muc_{31}}\nonumber\\
		&\times \Bigg\{
		\Bigg[ \muc_1 \Bigg( \hw_2^5 - \hw_3^5
		- \frac{\sin{\frac{5\pi}{b^2}}}{\sin{\frac{\pi}{b^2}}} \hw_2 \hw_3 (\hw_2^3 - \hw_3^3)
		+ 2 \cos{\textstyle\frac{2\pi}{b^2}}\ \frac{\sin{\frac{5\pi}{b^2}}}{\sin{\frac{\pi}{b^2}}}
		\hw_2^2 \hw_3^2 (\hw_2- \hw_3)\nonumber\\
		&-\frac{16 b^2}{\pi^2} \muc^{\frac{1}{b^2}} \cos^2{\textstyle\frac{\pi}{b^2}} \(
		\(1+4 \cos^2{\textstyle \frac{2\pi}{b^2}}\)(\hw_2^3-\hw_3^3)
		-\frac{\sin{\frac{3\pi}{b^2}}\sin{\frac{5\pi}{b^2}}}{\sin^2{\frac{\pi}{b^2}}}
		\hw_2 \hw_3 (\hw_2 - \hw_3)\)\nonumber\\
		&+\frac{1024 b^4}{\pi^4} \muc^{\frac{2}{b^2}} \cos^4{\textstyle \frac{\pi}{b^2}} 
		\cos^2{\textstyle \frac{2\pi}{b^2}} (\hw_2-\hw_3)
		\Bigg) - (1 \leftrightarrow 2) \Bigg]
		+\muc_3  \Bigg( \hw_1^5 - \hw_2^5\\
		&+ \frac{\sin{\frac{5\pi}{b^2}}}{\sin{\frac{\pi}{b^2}}}
		( \hw_3^4 (\hw_1 - \hw_2) - \hw_3 (\hw_1^4 - \hw_2^4) )
		- 2 \cos{\textstyle\frac{2\pi}{b^2}}\ \frac{\sin{\frac{5\pi}{b^2}}}{\sin{\frac{\pi}{b^2}}}
		( \hw_3^3 (\hw_1^2 - \hw_2^2) - \hw_3^2 (\hw_1^3 - \hw_2^3) )\nonumber\\
		&-\frac{16 b^2}{\pi^2} \muc^{\frac{1}{b^2}} \cos^2{\textstyle\frac{\pi}{b^2}}\ \Big(
		\(1+4 \cos^2{\textstyle \frac{2\pi}{b^2}}\)(\hw_1^3-\hw_2^3)
		+\frac{\sin{\frac{3\pi}{b^2}}\sin{\frac{5\pi}{b^2}}}{\sin^2{\frac{\pi}{b^2}}}
		( \hw_3^2 (\hw_1 - \hw_2) - \hw_3 (\hw_1^2 - \hw_2^2) )\Big)\nonumber\\
		&+\frac{1024 b^4}{\pi^4} \muc^{\frac{2}{b^2}} \cos^4{\textstyle \frac{\pi}{b^2}} 
		\cos^2{\textstyle \frac{2\pi}{b^2}} (\hw_1-\hw_2)
		\Bigg)
		\Bigg\}\nonumber
\end{align}
\begin{align}
	\hat{C}&(\beta_1,\beta_1,\beta_2|\tau_1,\tau_2,\tau_3)
		= \frac{a_2}{\muc_{12}\ \muc_{23}\ \muc_{31}}
		\Bigg\{\Bigg[ \muc_3  \Bigg( \hw_1^5 - 2 \cos{\textstyle\frac{2\pi}{b^2}}\ 
		\frac{\sin{\frac{3\pi}{b^2}}}{\sin{\frac{\pi}{b^2}}}\hw_2^5
		 + \frac{\sin{\frac{3\pi}{b^2}} \sin{\frac{5\pi}{b^2}}}{\sin^2{\frac{\pi}{b^2}}}
		(\hw_2^4 (\hw_3+\hw_1))
		\nonumber\\
		& - \frac{\sin{\frac{5\pi}{b^2}}}{\sin{\frac{\pi}{b^2}}} \hw_3 \hw_1^4
		- 2 \cos{\textstyle\frac{2\pi}{b^2}}\ \frac{\sin{\frac{5\pi}{b^2}}}{\sin{\frac{\pi}{b^2}}}
		\(\hw_2^3 \(\hw_3^2 + \hw_1^2 
		+ 4 \cos^2{\textstyle\frac{\pi}{b^2}} \hw_3 \hw_1\) - \hw_3^2 \hw_1^3\)
		\nonumber\\
		&- 2 \cos{\textstyle\frac{2\pi}{b^2}}\ 
		\frac{\sin{\frac{3\pi}{b^2}}\sin{\frac{5\pi}{b^2}}}{\sin^2{\frac{\pi}{b^2}}}
		\hw_3 \hw_1 \hw_2 (\hw_3 \hw_1 - \hw_1 \hw_2 - \hw_2 \hw_3)
		\nonumber\\
		&-\frac{16 b^2}{\pi^2} \muc^{\frac{1}{b^2}} \cos^2{\textstyle\frac{\pi}{b^2}} \Big(
		\(1+4 \cos^2{\textstyle \frac{2\pi}{b^2}}\)\hw_1^3
		- 2 \cos{\textstyle \frac{2\pi}{b^2}} \(1+4\cos{\textstyle \frac{2\pi}{b^2}}\) \hw_2^3
		- \frac{\sin{\frac{5\pi}{b^2}}}{\sin{\frac{\pi}{b^2}}} \hw_3 \hw_1^2
		\nonumber\\
		&+\frac{\sin{\frac{3\pi}{b^2}}\sin{\frac{5\pi}{b^2}}}{\sin^2{\frac{\pi}{b^2}}}
		\hw_2^2 (\hw_3 + \hw_1)
		- 2 \cos{\textstyle\frac{2\pi}{b^2}}\ \frac{\sin{\frac{5\pi}{b^2}}}{\sin{\frac{\pi}{b^2}}}
		(\hw_2 (\hw_3^2 + \hw_1^2 + \hw_3 \hw_1) - \hw_3^2 \hw_1 )
		\Big)\nonumber\\
		&+\frac{1024 b^4}{\pi^4} \muc^{\frac{2}{b^2}} \cos^4{\textstyle \frac{\pi}{b^2}} 
		\cos^2{\textstyle \frac{2\pi}{b^2}} (\hw_1-\hw_2)
		\Bigg) - (3 \leftrightarrow 1) \Bigg]\\
		&+\muc_2 \Bigg( \hw_3^5 - \hw_1^5
		- \frac{\sin{\frac{5\pi}{b^2}}}{\sin{\frac{\pi}{b^2}}} \hw_3 \hw_1 (\hw_3^3 - \hw_1^3)
		+ 2 \cos{\textstyle\frac{2\pi}{b^2}}\ \frac{\sin{\frac{5\pi}{b^2}}}{\sin{\frac{\pi}{b^2}}}
		\hw_3^2 \hw_1^2 (\hw_3- \hw_1)\nonumber\\
		&-\frac{16 b^2}{\pi^2} \muc^{\frac{1}{b^2}} \cos^2{\textstyle\frac{\pi}{b^2}} \(
		\(1+4 \cos^2{\textstyle \frac{2\pi}{b^2}}\)(\hw_3^3-\hw_1^3)
		-\frac{\sin{\frac{3\pi}{b^2}}\sin{\frac{5\pi}{b^2}}}{\sin^2{\frac{\pi}{b^2}}}
		\hw_3 \hw_1 (\hw_3 - \hw_1)\)\nonumber\\
		&+\frac{1024 b^4}{\pi^4} \muc^{\frac{2}{b^2}} \cos^4{\textstyle \frac{\pi}{b^2}}
		 \cos^2{\textstyle \frac{2\pi}{b^2}} (\hw_3-\hw_1)
		\Bigg)\Bigg\}\nonumber
\end{align}
All of the above results satisfy the required properties of being real and cyclically symmetric in the momenta. In particular, cyclic symmetry is evident for $(n_1,n_2,n_3)=(0,0,0)$ and $(1,1,1)$, while we checked it for the other cases explicitly (notice that in order to check the symmetry one also has to interchange the chiralities of the operators and thus to take into account the fact that different renormalization factors must be used).

\subsection{$d$ from $C$ once again}
\label{aptwothree}

Our purpose here is to do a consistency check, namely to verify that
the $c=1$ limit of the three-point function 
produces the same two-point function which was derived in section \ref{bndtwopsec}.
In fact, we can do better and check that one function follows from the other
{\it before} taking the limit, but after the momenta degenerate in the matter sector 
were substituted and all special functions and integrals were eliminated. 
Then the agreement of the limiting expressions trivially follows.

In other words, we are going to show that if one takes the expression \eqref{3ptfullfinal},
puts $(n_1,n_2,n_3)=(0,n,n)$, $(\tau_1,\tau_2,\tau_3)=(\tau,\tau,\tau')$ and
integrates with respect to $-\mubc(\tau)$, one obtains the result \eqref{Wtwobnd}.
Of course, one can choose two other possibilities to get the two-point correlator
similar to the ones in appendix \ref{aptwoother}. However, they require
more complicated calculations, and we proceed here only with the first choice.  

First, one can show that, after setting $n_1=0$, $n_2=n_3=n$, 
\eqref{3ptfullfinal} becomes
\begin{equation}\label{3ptfulln}
\hat C(b,\beta_{n},\beta_{n}|\tau_1,\tau_2,\tau_3) 
= - a_n \Bigg\{
\frac{\scriptstyle \prod\limits_{k=-n}^{n}  \left(
\hat W(\tau_2)-\hat W(\tau_3-2\pi i k)\right)}
{\hat\mu_{12}\,\hat\mu_{23}}
- \frac{\scriptstyle \prod\limits_{k=-n}^{n} \left(
\hat W(\tau_1)-\hat W(\tau_3-2\pi i k)\right)}
{\hat\mu_{12}\,\hat\mu_{13}}
\Bigg\},
\end{equation}
where the coefficient $a_n$ is defined in \eqref{const_c}.
Then, taking the limit $\tau_1=\tau_2=\tau$, $\tau_3=\tau'$, one obtains
\be
\hat C(b,\beta_{n},\beta_{n}|\tau,\tau,\tau') 
= - {a_n \over \hat\mu_{\tau\tau'}}\left\{ 
\frac{\scriptstyle \prod\limits_{k=-n}^{n}  \left(
\hat W(\tau)-\hat W(\tau'-2\pi i k)\right)}
{\hat\mu_{\tau\tau'}} 
- \frac{{\p_\tau}\scriptstyle \prod\limits_{k=-n}^{n}  \left(
\hat W(\tau)-\hat W(\tau'-2\pi i k)\right)}
{{\p_\tau}\hat\mu_{\tau\tau'}}
\right\}.
\label{3pttaun}
\ee
The integral is easily evaluated producing
\be
-\int d\mubc(\tau) \, \hat C(b,\beta_{n},\beta_{n}|\tau,\tau,\tau') 
= - {a_n \over \hat\mu_{\tau\tau'}}\prod\limits_{k=-n}^{n}  \left(
\hat W(\tau)-\hat W(\tau'-2\pi i k)\right).
\label{twothreeres}
\ee
This result coincides with the two-point function $\hat d(\beta_n|\tau,\tau')$ 
from \eqref{Wtwobnd}.
This constitutes a check of the correctness of our intermediate calculations
in the derivation of the three-point function.

\addcontentsline{toc}{section}{References}
\bibliographystyle{../../bibtex/myutcaps}
\bibliography{../../bibtex/mybib}

\end{document}